\documentclass[]{aastex631}

\usepackage{graphicx} 
\usepackage{ulem}
\usepackage{subfigure}

\newcommand{\HH}{H$_2$}
\newcommand{\HHO}{H$_2$O}
\newcommand{\OO}{O$_2$}
\newcommand{\CHHHH}{CH$_4$}
\newcommand{\COO}{CO$_2$}
\newcommand{\CO}{CO}

\newcommand{\He}{He}
\newcommand{\Fe}{Fe}
\newcommand{\FeO}{FeO}
\newcommand{\FeFeOOO}{Fe$_2$O$_3$}

\shortauthors{Seo, Ito, \& Fujii}
\usepackage{lipsum}
\usepackage{float}
\usepackage{comment}
\usepackage{tabularx}
\usepackage{mathtools}
\usepackage[flushleft]{threeparttable}

\begin{document}

\title{Role of magma oceans in controlling carbon and oxygen of sub-Neptune atmospheres}

\author[0000-0002-0749-2090]{Chanoul Seo}
\affiliation{Department of Astronomical Science, School of Physical Sciences, Graduate University for Advanced Studies (SOKENDAI), 2-21-1 Osawa, Mitaka, Tokyo 181-8588, Japan}
\affiliation{Division of Science, National Astronomical Observatory of Japan, 2-21-1 Osawa, Mitaka, Tokyo 181-8588, Japan}

\author[0000-0002-0598-3021]{Yuichi Ito}
\affiliation{Division of Science, National Astronomical Observatory of Japan, 2-21-1 Osawa, Mitaka, Tokyo 181-8588, Japan}
\affiliation{Faculty of Science and Technology,
Sophia University,
Kioi-Cho 7-1, Chiyoda-ku, Tokyo, 102-8554 Japan}
\affiliation{Department of Physics and Astronomy, University, College London 
Gower Street, WC1E 6BT London, United Kingdom}

\author[0000-0002-2786-0786]{Yuka Fujii}
\affiliation{Department of Astronomical Science, School of Physical Sciences, Graduate University for Advanced Studies (SOKENDAI), 2-21-1 Osawa, Mitaka, Tokyo 181-8588, Japan}
\affiliation{Division of Science, National Astronomical Observatory of Japan, 2-21-1 Osawa, Mitaka, Tokyo 181-8588, Japan}

\correspondingauthor{Chanoul Seo}
\email{chanoul.seo@grad.nao.ac.jp}

\begin{abstract}
Most exoplanets with a few Earth radii are more inflated than bare-rock planets with the same mass, indicating a substantial volatile amount. Neither the origin of the volatiles nor the planet's bulk composition can be constrained from the mass-radius relation alone, and the spectral characterization of their atmospheres is needed to solve this degeneracy. Previous studies showed that chemical interaction between accreted volatile and possible molten rocky surface (i.e., magma ocean) can greatly affects the atmospheric composition. However, a variety in the atmospheric compositions of such planets with different properties remains elusive. In this work, we examine the dependence of atmospheric H, O, and C on planetary parameters (atmospheric thickness, planetary mass, equilibrium temperature, and magma properties such as redox state) assuming nebula gas accretion on an Earth-like core, using an atmosphere-magma chemical equilibrium model. Consistent with previous work, we show that atmospheric \HHO\ fraction on a fully molten rocky interior with an Earth-like redox state is on the order of $10^{-2}$-$10^{-1}$ regardless of other planetary parameters. Despite the solubility difference between H- and C-bearing species, C/H increases only a few times above the nebula value except for atmospheric pressure $\lesssim$1000 bar and \HHO\ fraction $\gtrsim$10\%. This results in a negative O/H-C/O trend and depleted C/O below one-tenth of the nebula gas value under an oxidized atmosphere, which could provide a piece of evidence of rocky interior and endogenic water. We also highlight the importance of constraints on the high-pressure material properties for interpreting the magma-atmospheric interaction of inflated planets.
\end{abstract}


\section{Introduction}

Past exoplanet surveys have revealed that exoplanets within the radius range spanning Earth to Neptune-sized bodies are abundant \cite[e.g.,][]{2010Sci...330..653H, 2012ApJS..201...15H, 2013ApJ...766...81F, 2019AJ....158..109H}. While some have a mass-radius relationship consistent with the bulk composition similar to that of Earth, others have larger radii than those of bare Earth-like planets with the same masses, suggesting that they contain thick layers of volatile species \cite[e.g.,][]{2014ApJ...783L...6W, 2015ApJ...801...41R, 2019AREPS..47...67O}. The former tend to be called super-Earths, while the latter are often called sub-Neptunes, although the exact definition of these terms can depend on the context.

The origin of these intermediate-sized planets close to their host stars has been the subject of active debate. One hypothesis posits that super-Earths and sub-Neptunes share a common origin characterized by a rocky core enveloped by an H-He atmosphere accreted from the protoplanetary disk and finds that such a model can explain the distinctive radius gap in observed exoplanets' radii at approximately $1.8R_{\oplus}$ \cite[e.g.,][]{2017AJ....154..109F, 2018MNRAS.479.4786V, 2022AJ....163..179P}. Previous works have discussed the possible origins of this radius gap, including atmospheric escape, impact loss, and accretion processes \cite[e.g.,][]{2019AREPS..47...67O, 2020MNRAS.491..782W, 2022ApJ...941..186L}. Another hypothesis proposes that the planets with larger radii accreted a large amount of icy planetesimals beyond the so-called snow lines \cite[e.g.,][]{2019PNAS..116.9723Z, 2020ApJ...896L..22M, 2020A&A...643L...1V, 2021ApJ...923..247Z, 2022Sci...377.1211L, 2024NatAs...8..463B}. An examination of these proposed scenarios for the formation of such close-in small planets can be found in the review by \citet[]{2021JGRE..12606639B}, along with pertinent references contained therein. 

Given that the mass-radius relationship alone cannot uniquely constrain the bulk composition \citep[e.g.,][]{2010ApJ...716.1208R, 2013ApJ...775...10V, 2023ApJ...947L..19R}, more clues to the origins of sub-Neptunes are expected to be obtained through atmospheric spectroscopy. This is particularly true as the James Webb Space Telescope (\textit{JWST}) data become available with higher precision than past observations and start to constrain the atmospheric compositions (typically at the 0.01-1~bar levels) of sub-Neptunes/super-Earths (e.g., \citealt{2023ApJ...948L..11M} for GJ 486b; \citealt{2023ApJ...956L..13M} for K2-18b;  \citealt{2023arXiv231010711M} for GJ 1132b; \citealt{2024arXiv240504744H} for 55 Cancri e; \citealt{2024arXiv240303325B} for TOI-270 d). 

When interpreting these atmospheric data, it is important to note that the atmospheric composition of these small planets could result from interactions between accreted volatile species and the planetary core. Therefore, studying how the accreted volatiles are altered and partitioned within the planet is critical \cite[e.g.,][]{2012E&PSL.341...48H, 2020GeCoA.280..281G, 2020SciA....6.1387S, 2020NatCo..11.2007D, 2022E&PSL.57717255G}. Several review papers \citep[e.g.,][]{2021SSRv..217...22G, 2022ARA&A..60..159W, 2023FrEaS..1159412S, 2023ASPC..534..907L, 2023SSRv..219...51S, 2024arXiv240504057L} provide a comprehensive survey of the previous studies on the impacts of core-atmospheric chemistry on terrestrial planet atmospheres, starting with the solar system planets—Earth, Venus, and Mars—and more recently extending to larger rocky exoplanets such as super-Earths and sub-Neptunes.

One of the new insights in the context of exoplanets is that the magma (and iron core) of sub-Neptunes serve as crucial volatile reservoirs, especially for hydrogen, through the dissolution of significant amounts of \HH\ and \HHO. If the total amount of volatiles is within a few percent, this role of the planetary core as a volatile reservoir significantly influences the planet's evolution and causes a decoupling between the amount of volatiles in the interior and the planetary radius \cite[e.g.,][]{2018ApJ...854...21C, 2019ApJ...887L..33K, 2020ApJ...891..111K, 2022PSJ.....3..127S}. In addition to the radius-volatile content decoupling effect, the ingassing of certain volatile species affects the atmospheric composition by changing the elemental abundances of the atmosphere. For instance, \citet{2024arXiv240105864S} used an atmosphere-magma model of K2-18b to demonstrate that the high solubility of nitrogen in magma in a reducing environment can deplete N from the atmosphere depending on the redox state. Based on this result, they argued that using N depletion as the sole indicator for the existence of a water ocean can be misleading.

Another process through which the rocky interior affects the atmospheric composition is the redox reactions between the atmosphere and the magma. Notably, \HH, the major atmospheric species of sub-Neptunes, can be oxidized by magma to produce substantial amounts of \HHO\ \cite[e.g.,][]{1990orea.book..195S, 2006ApJ...648..696I}. \citet{2020ApJ...891..111K} showed that for sub-Neptunes with relatively thin atmospheres, such alteration of atmospheric composition by magma challenges the interpretation of the atmospheric composition, as a certain \HHO\ mixing ratio can be explained by different combinations of the accreted  volatile composition and the redox state of magma. \citet{2021ApJ...909L..22K} proposed that water-rich atmospheres originated from the interaction between the primordial nebula-driven gas and magma would have the C/O ratio smaller than the value expected from the ice solids, which can be interpreted as the evidence of magma-atmosphere reaction. The C/O ratio change due to magma was also recently discussed by \citet{2024ApJ...963..157T} using their magma-atmosphere H-C-N-O-S chemical reaction model. They argued that the C/O ratio is difficult to use as an indicator of the formation history of rocky planets due to its susceptibility to internal chemistry variations. Another parameter that can imply the presence of magma is the \COO/\CO\ ratio. \citet{2024arXiv240105864S} suggested that a small \COO/\CO\ ratio in the deep atmosphere of K2-18b can indicate the presence of high-temperature magma that reacts with the atmosphere. However, the \COO/\CO\ ratio at the altitudes observed through the transmission spectroscopy can differ from that in the deep atmosphere, and the former varies depending on the elemental abundance of the atmosphere and the efficiency of the vertical mixing rather than directly constraining the presence of magma.

While these previous works pointed out the importance of chemical interactions between the atmosphere and the planetary core in determining the final state, they did not explicitly examine how the final atmospheric composition depends on assumptions about planetary parameters. Several previous studies \citep[e.g.,][]{2024ApJ...963..157T, 2024arXiv240105864S, 2024arXiv240509284M} have calculated the atmospheric composition of the H-C-N-O-(S) elemental system by assuming the final redox state of the magma. However, the final redox state of the system fundamentally depends on the relative amount of reactive magma compared to the atmosphere, the composition of the magma (iron content), and the iron speciation prior to the chemical reaction. These are parameters that are also relevant to planet formation scenarios. Although these magma properties have been considered in some existing studies \citep[e.g.,][]{2020ApJ...891..111K, 2021ApJ...909L..22K}, quantitative analyses for a wider range of elements (such as Carbon) are limited. Given the accumulating data from atmospheric spectroscopy of super-Earths and sub-Neptunes, it is also important to investigate the trends of the main atmospheric components, represented by O/H and C/O, as functions of various observable planetary parameters related to atmospheric structure (e.g., planetary radius, equilibrium temperature, mass).

In this paper, we study such trends based on chemical-equilibrium calculations between the atmosphere and underlying magma, focusing on the scenario where sub-Neptunes form through the accretion of nebula gas onto rocky cores interior to the \HHO\ snowline \citep{2017ApJ...847...29O, 2023ApJ...943...11M}. We show specific relations between O/H and C/O and their dependencies on magma properties (effective amount, Fe fraction, and initial Fe speciation), observable planetary parameters (planetary mass, equilibrium temperature, and radius), and an additional parameter for determining atmospheric structure (Radiative-Convective Boundary). Furthermore, we introduce the concept of planetary evolution using a toy model and discuss the impact of magma solidification on atmospheric composition. We explore how the O/H-C/O relation and their ranges could potentially help to distinguish the origins of sub-Neptunes from other major formation scenarios, such as the assembly of icy planetesimals. In addition to the numerical calculations with the non-ideal behavior of major species, analytic expressions will be presented for future use. 

In section 2, we introduce the structure, chemical reactions, and other assumptions of our sub-Neptune model. Section 3 presents the main results, showing the range of atmospheric compositions and their dependence on the planetary parameters. In section 4, we will address the implications of our findings on the formation scenario, potential spectral features, and the impact of different atmospheric structure models and atmospheric escape mechanisms on atmospheric composition. Section 5 summarizes our discussion.

\section{Model}
\label{s:method}

Our study aims to investigate atmospheric compositions focusing on H-, O-, and C-bearing species under the formation scenario where the nebula gas accretes onto a rocky core \cite[e.g.,][]{2017ApJ...847...29O}. The structure of our model follows that of \citet{2020ApJ...891..111K}. However, we also introduce several modifications to their model. Specifically, we incorporate C-bearing species into the analysis and derive the surface conditions (temperature and pressure) from the atmospheric temperature profile and observable planetary parameters such as planetary mass, radius, and equilibrium temperature derived from stellar irradiation without using the pre-assumed value. We also introduce an additional buffer system, the Ferric-Ferrous equilibrium, alongside the Iron-W{\"u}stite buffer to expand the range of affordable redox states towards a more oxidized regime. 

The model components and assumptions are described in depth in this section.

\subsection{Vertical structure}
\label{ss:model_structure}

\begin{figure}
\centering
\includegraphics[width=8.5cm]{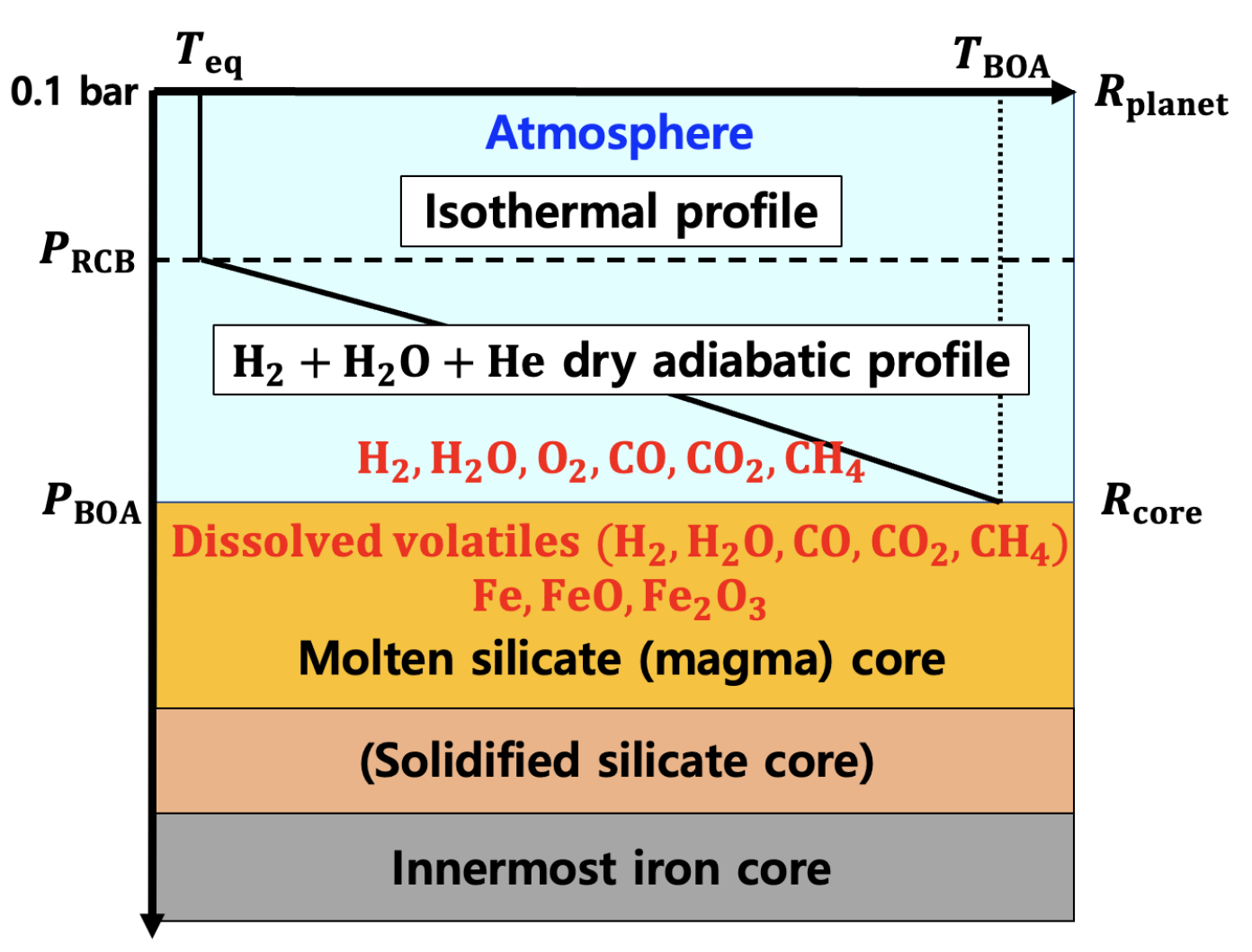}
\caption[Model structure]{The schematic figure of the vertical structure of the sub-Neptune model presented in this study. The black line within the atmosphere layer illustrates the assumed temperature-pressure relationship in our model. Chemical species involved in internal reactions are written in red within each structural layer. The RCB stands for the Radiative-Convective Boundary, while the BOA denotes the Bottom of the Atmosphere.} 
\label{fig:schematics_vertical}
\end{figure}

Our planetary model consists of three layers: an atmosphere, a silicate layer, and an innermost iron core, as shown in Figure \ref{fig:schematics_vertical}. 

We assume a vertically homogeneous composition in the atmosphere when calculating the bulk atmospheric composition for simplicity. The elemental ratio of helium to hydrogen in the atmosphere (He/H) is fixed at 0.1, reflecting the nebula gas value (0.079, from \citealt{2003ApJ...591.1220L}). This assumption implies that the partitioning of \HH\ to the magma is comparable to that of He, which could be supported by the similar mean molecular radius between He and \HH\ \cite[e.g.,][]{2012E&PSL.345...38H,2023FrEaS..1159412S}. We note that there remains uncertainties in this assumption because the low-pressure experimental study shows smaller solubility of He than \HH\ \cite[e.g.,][]{1986GeCoA..50..401J}. However, since the solubility of He does not significantly affect our results, we assume a fixed nebula gas-like atmospheric He/H ratio for simplicity. Only when we demonstrate the transmission spectra in Section \ref{ss:implicationobs} do we model the vertical profile of atmospheric molecules considering atmospheric chemistry and eddy diffusion (considered indirectly by assuming specific quenching points.)

As the fiducial model, the atmospheric thermal profile is assumed to be composed of an isothermal stratosphere with equilibrium temperature, $T_\mathrm{eq}$, above the radiative-convective boundary (RCB) and the dry adiabatic troposphere between the RCB and the bottom of the atmosphere (BOA). The planetary radius, $R_{\rm p}$, is set at the pressure level of 0.1 bar, consistent with the altitude of the optical radius of planets \citep[e.g.,][]{2018ApJ...853....7K}. The pressure of the RCB, $P_{\rm RCB}$, is treated as a variable and is either 0.1, 1, 10, or 100 bars. This simplified Temperature-Pressure profile allows for easy analytic estimates and helps us clarify the dependence on planetary parameters in Section \ref{ss:atmospheric composition_planetaryparamdepen}. We will also discuss the difference between our simplified thermal profile and a more realistic radiative-convective equilibrium profile in Section \ref{ss:discussion_TP}. 

In our model, the atmosphere is assumed to be in hydrostatic equilibrium and divided into 1000 layers, evenly spaced in altitude between the RCB and BOA, and 100 layers evenly spaced in pressure between the $R_{\rm p}$ and the RCB. With this layered setup, we account for the altitude dependence of the gravitational acceleration $g$ and the temperature dependence of the heat capacity $C_p$ when solving for structure assuming hydrostatic equilibrium. The lapse rate in the lower dry adiabatic troposphere is computed from $g/C_p$, and for $C_p$, we take into account the temperature dependence of the $C_p$ of \HH, \HHO, and He, which account for over 95\% of the atmosphere in our results. We refer to the JANAF thermochemical table (\citealt{219851}) for the $C_p$ of the three species, which covers the temperature range from 100~K to 6000~K. 

The innermost iron core is assumed to take up 34\% of the planetary mass, consistent with Earth's iron core mass fraction, and the remaining 66\% of the planetary mass is attributed to the silicate layer. We ignore the atmospheric mass in this study considering the dominant mass of the planetary core. 
We calculate the radius of the planetary (silicate + iron) core using the mass-radius relationship equations of rocky cores from \citet{2007ApJ...659.1661F} (see their Figure 4 and Equation 8). Note that our study currently omits the chemical impact of the iron core. This assumption would align with scenarios where the segregation of the iron core occurred early, preventing the iron core from being involved in reactions with magma. The efficacy of iron core segregation depends on various factors, such as heat flow and iron droplet size \citep[e.g.,][]{2021ApJ...914L...4L}. 
In cases of sufficiently vigorous magma convection, it has been argued that the formation of the iron core may be delayed, allowing sufficient time for the iron core (and drifting iron droplets) to attain equilibrium with the surrounding magma and atmosphere \cite[e.g.,][]{2021ApJ...914L...4L}. 
The potential impact of the innermost iron core on volatile partitioning is described in Section \ref{ss:Limitation}.

The silicate layer is stratified into an upper molten magma and a lower solidified rock layer. Under nominal conditions, magma convection is assumed vigorous enough to ensure homogeneous mixing and fully reactive magma. The dependence of atmospheric composition on the reactive amount of magma is discussed in Section \ref{ss:atmospheric composition_magmaparamdepen}.

In the fiducial model, we assume the elemental composition of the magma to be the same as the Mid Ocean Ridge Basalt (MORB), with elemental ratios of $\rm Fe/Si = 0.17$ and $\rm Mg/Si = 0.22$ \citep[and references therein]{2013GGG....14..489G}. We note that this MORB-like composition differs from the peridotite-like Bulk Silicate Earth (BSE), which is characterized by a relatively higher magnesium fraction ($\rm Mg/Si \sim 0.92$, \citealt{2010E&PSL.293..259J}). This compositional difference can influence the equilibrium state by altering the abundance of reactive species (e.g., FeO) and the solubility of gases such as \HH, \HHO, and \COO\ \cite[e.g.,][]{2012E&PSL.345...38H, 2022CoMP..177...40A}. We explore the dependence of atmospheric composition on the Fe/Si ratio in Section \ref{ss:atmospheric composition_magmaparamdepen} and discuss the potential effects of varying gas solubility (\HH, \HHO, and \COO) in Section \ref{ss:Limitation}.

The lower solidified rock layer remains chemically inert and is assumed to be free of volatiles. We presume the molten fraction of the silicate layer to be 100\% and introduce solidified magma layers into the model when discussing the magma solidification effect. Section \ref{ss:magma_solidification} describes in detail how we account for magma solidification in this study.

Importantly, we calculate the chemical equilibrium state (Section \ref{ss:model_redox_chemistry} and Section \ref{ss:model_dissolution} below) using the temperature and pressure conditions at the interface between the magma and the atmosphere. In other words, we assume that chemical reactions occur at the magma surface and the lowest atmospheric layer. In reality, the equilibrium state can be dependent on the pressure and the saturation amount of each volatile species that can vary with depth in the magma. Such effect will be the scope of the future work. 

\subsection{Redox reactions}
\label{ss:model_redox_chemistry}

In our chemical equilibrium calculation, we consider \HH, \HHO, \OO, \CO, \COO, \CHHHH, \Fe, \FeO, and \FeFeOOO\ as participating species. 
Note that \He\ is one of the major gas components determining the atmospheric profile but does not participate in the chemical reactions. Its abundance remains fixed at the nebula gas value (He/H = 0.1), as mentioned in Section \ref{ss:model_structure}. 

We consider the redox equilibrium among these species as follows \citep[see also][]{2022PSJ.....3..127S}. 

In the atmosphere, 
\begin{itemize}
    \item 2 \HH\ + \OO\ $\leftrightarrows $ 2 \HHO,
    \item 2 \CO\ + \OO\ $\leftrightarrows $ 2 \COO,
    \item \CHHHH\ + \HHO\ $\leftrightarrows $ \CO\ + 3 \HH. 
\end{itemize}

In the molten silicate layer, we consider the Iron-W\"{u}stite (IW) reaction (as assumed in \citealt{2020ApJ...891..111K}) and the Ferric-Ferrous reaction (in the most oxidized case) as the redox buffers:
%%%%%%%%%%
\begin{itemize}
    \item 2 Fe\ + \OO\ $\leftrightarrows $ 2 \FeO
    \item 2 FeO\ + 0.5 \OO\ $\leftrightarrows $ \FeFeOOO
\end{itemize}
%%%%%%%%%%

We include the non-ideal behavior of atmospheric species by utilizing the fugacity ($f\rm S$, where ${\rm S}$ represents each species) instead of partial pressure in our equilibrium calculations. The equilibrium between the molten rock and the atmosphere indicates that the fugacity of \OO\ ($f{\rm O_2}$) in these equations is equated. The fugacity of the species $f\rm S$ is determined by $f{\rm S}=\phi_{\rm S} P_{\rm S}$, where $\phi_{\rm S}$ is the fugacity coefficient and $P_{\rm S}$ is the partial pressure. We calculate the fugacity coefficient of gases from the Equation of State (EoS) formulation by \citet{2009GeCoA..73.2089Z}.

For reactions between fluids, the equilibrium constant is determined using the Gibbs free energy:
\begin{equation}
K_p = \exp \left( -\frac{\Delta G^{\circ }}{RT } \right)
\end{equation}
where $K_{\rm p}$ denotes the equilibrium constant,  $\Delta G^{\circ }$ is the difference in standard Gibbs free energy of formation between the reactant(s) and the product(s), $R$ is the gas constant, and $T$ is the temperature. 
The standard Gibbs free energy of formation for volatile species is again adopted from the JANAF table (\citealt{219851}).

For the reactions in the molten silicate layer, we include the change in volume of the reaction ($\Delta V$) term:
\begin{equation}
K_p = \exp \left(-\frac{\Delta G^{\circ }+\int_{P_{0}}^{P} \Delta VdP}{RT}\right).
\end{equation}

The $\Delta G^{\circ }$ of the IW buffer is obtained from \citet{kowalski1995thermodynamic}, which is validated for temperatures up to 3000~K. We refer to \citet{2014JGRB..119.4164K} for the EoS of liquid Fe, validated up to 3300~K and $10^{6}$~bar. For the EoS of liquid FeO, we utilize the data from \citet{2019Sci...365..903A}, which is validated under conditions up to 3000~K and $3\times10^{5}$~bar. We consider the non-ideal behavior of liquid FeO by introducing the activity coefficient of 1.5 (\citealt{1997ChGeo.139...21H, 2020ApJ...891..111K}). On the other hand, we assume that liquid Fe remains in a pure state within the magma, corresponding to an activity coefficient of 1. 

In the case of the Ferric-Ferrous equilibrium, we adopt the equilibrium constant equation proposed by \citet{2022GeCoA.328..221H}. 
This equation incorporates the EoS of liquid Fe from \citet{2014JGRB..119.4164K} and \FeFeOOO\ from \citet{2019Sci...365..903A} and \citet{2020NatCo..11.2007D}. 
Additionally, it includes a term describing the dependence of the activity coefficients of \FeO\ and \FeFeOOO\ on the silicate composition based on multiple experimental studies rather than assuming a fixed value. The suggested equation from \citet{2022GeCoA.328..221H} has the form of
\begin{equation}
\textrm{log} \left(\frac{X_{\textrm{FeO}_{1.5}}}{X_{\textrm{FeO}}}\right) = a\textrm{log}f\textrm{O}_{2}-\frac{\left(\Delta G^{\circ }+\int_{P_{0}}^{P} \Delta VdP \right)}{{RT\textrm{ln}(10)}}-\textrm{log}\left(\frac{\gamma_{\textrm{FeO}_{1.5}}}{\gamma_{\textrm{FeO}}}\right)
\end{equation}
where $a$ is a stoichiometric coefficient and $\gamma$ is the activity coefficient, which is the function of temperature and silicate composition. In \citet{2022GeCoA.328..221H}, $a$ has an empirical value of 0.1917. We note that this is different from the $a=0.25$ of the ideal case, which we can expect from the reaction (FeO\ + 0.25 \OO\ $\leftrightarrows $ FeO$_{1.5}$). The validation range of temperature and pressure for this equation extends up to 2200~K and 70000~bar, respectively.

The $f\rm O_{2}$ parameterization in magma depends on the metal saturation in the magma ocean \cite[e.g.,][]{2022PSJ.....3..115I}. We assume the $[\textrm{Fe}]$, the molar fraction of iron in the metal phase, is always 1 (pure iron). The IW buffer has a limited range of $f\rm O_{2}$ it can cover. Under the reduced condition (low $f\rm O_{2}$), pure iron exists in the magma, and both the IW buffer and the Ferric-Ferrous equilibrium occur concurrently ($\textrm{Fe} \longleftrightarrow \textrm{FeO} \longleftrightarrow \textrm{Fe}_{2}\textrm{O}_{3}$). As the magma becomes more oxidized, the $\textrm{Fe}$ oxidizes to \FeO, and \FeO\ changes further to \FeFeOOO. Under the oxidized condition where the $f\rm O_{2}$ level exceeds the range covered by the IW buffer, pure iron becomes depleted, and the ferric-ferrous equilibrium solely determines the $f\rm O_{2}$ ($\textrm{FeO} \longleftrightarrow \textrm{Fe}_{2}\textrm{O}_{3}$). 

\begin{figure}
\centering
\includegraphics[width=18cm]{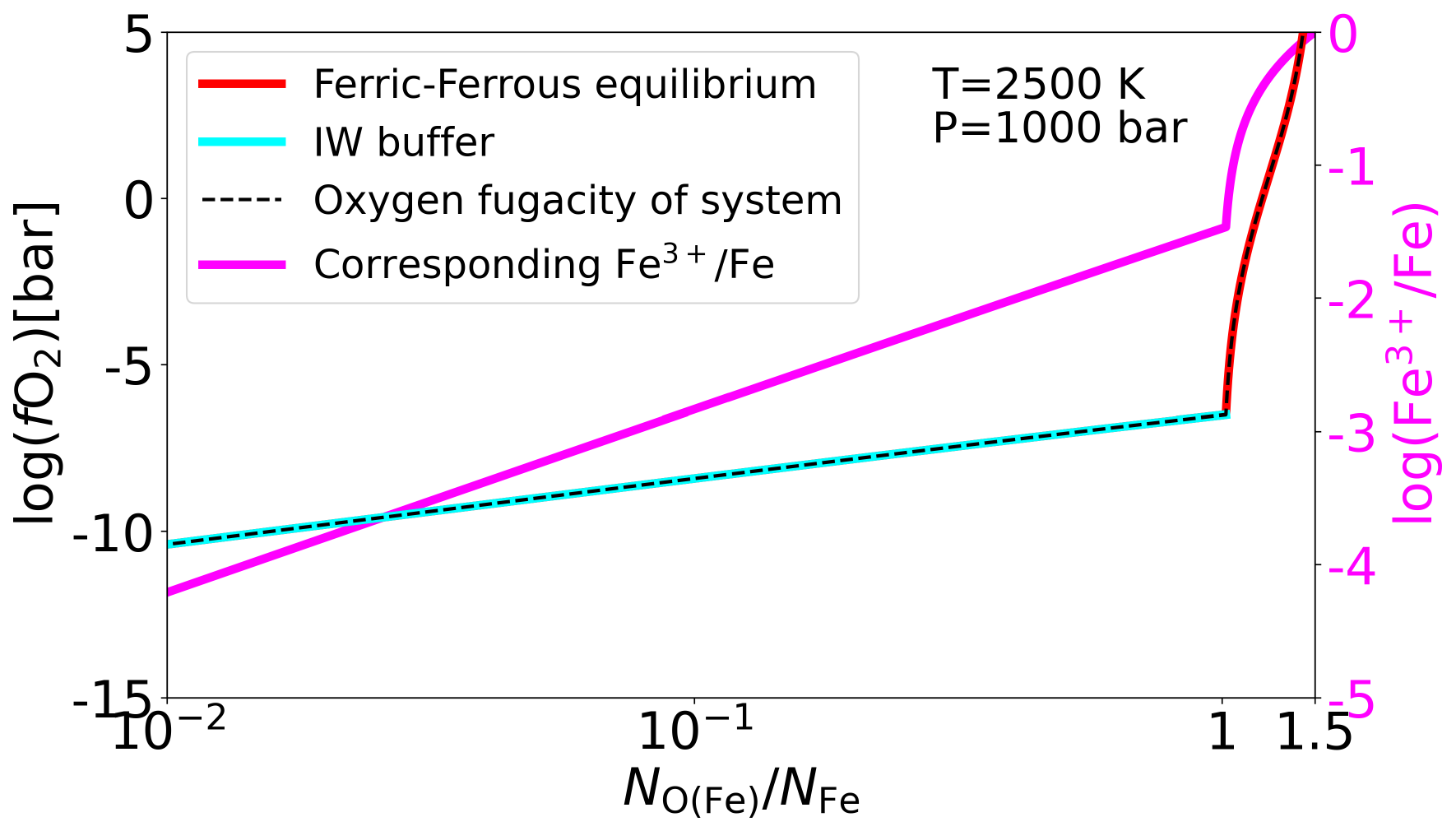}
\caption[Magma redox reactions]{The relation between the iron speciation and the oxygen fugacity. The x-axis is the number ratio between the oxygen bound to iron and iron itself, while the y-axis denotes the oxygen fugacity ($f\rm O_{2}$). The cyan (red) plot indicates the oxygen fugacity calculated from the IW buffer (Ferric-Ferrous equilibrium). The black dashed line represents the actual $f\rm O_{2}$ of the system. The magenta-colored line shows the $\rm Fe^{3+}/Fe$ ratio corresponding to each $N_{\rm O(Fe)}/N_{\rm Fe}$. The temperature and pressure are set at 2500 K and 1000 bar, respectively.}
\label{fig:FeFeOFe2O3relation}
\end{figure}

Figure \ref{fig:FeFeOFe2O3relation} illustrates the relationship between iron speciation and the oxygen fugacity of the system. Here, we parameterize the iron speciation of magma by the number ratio between the oxygen bound to iron and iron itself, denoted as $N_{\rm O(Fe)}$/$N_{\rm Fe}$. Depending on the oxidation state, $N_{\rm O(Fe)}$/$N_{\rm Fe}$ ranges from 0 (all iron exists as pure Fe) to 1.5 (all iron exists as \FeFeOOO). The figure clearly shows the O-rich magma corresponds to higher $f{\rm O_2}$. We utilize $N_{\rm O(Fe)}$/$N_{\rm Fe}$ as the primary parameter to indicate the redox state of the magma hereafter.

%%%%%%%%%%%%%%%%%%%%
%%%%%%%%%%%%%%%%%%%%

We note that the $f{\rm O_2}$ shows higher sensitivity to the $N_{\rm O(Fe)}$/$N_{\rm Fe}$ after the depletion of pure Fe due to the change in governing reactions and the specific parameterization, $N_{\rm O(Fe)}$/$N_{\rm Fe}$, that we use. From the chemical equations of IW buffer and Ferric-Ferrous equilibrium, the equilibrium constants for the two buffers are
\begin{eqnarray}
K_{\rm IW} =\frac{({\gamma_{\rm FeO}}\textrm{[FeO]})^{2}}{f{\rm O_2}}\\
K_{\rm FF} = \left(\frac{\gamma_{\rm FeO_{1.5}}[\rm FeO_{1.5}]}{\gamma_{\rm FeO}[\rm FeO]}\right)^{1/a}\frac{1}{f{\rm O_2}}
\end{eqnarray}
where $a$ represents the stoichiometric coefficient of \OO\ in the Ferric-Ferrous equilibrium, and $K_{\rm IW}$ and $K_{\rm FF}$ denote the equilibrium constants of the IW buffer and the Ferric-Ferrous equilibrium, respectively.  
From the two equilibrium constant equations, it is clear that the proportional relationship shifts from $\textrm{log}(f{\rm O_2}) \propto \rm log([\FeO])$ to $\textrm{log}(f{\rm O_2}) \propto \rm \frac{1}{a}log(\frac{[\rm FeO_{1.5}]}{[\rm FeO]})$ as the IW buffer is excluded from the reaction. 
Therefore, $f{\rm O_2}$ under oxidized conditions rapidly changes as $N_{\rm O(Fe)}$/$N_{\rm Fe}$ increases because the ratio between two iron species determines $f{\rm O_2}$.

While $f\rm O_{2}$ also is related to the amount of \OO, we do not track \OO. $f\rm O_{2}$ serves only as the variable that indicates the redox state of the system. 

\begin{figure}
\centering
\includegraphics[width=18cm]{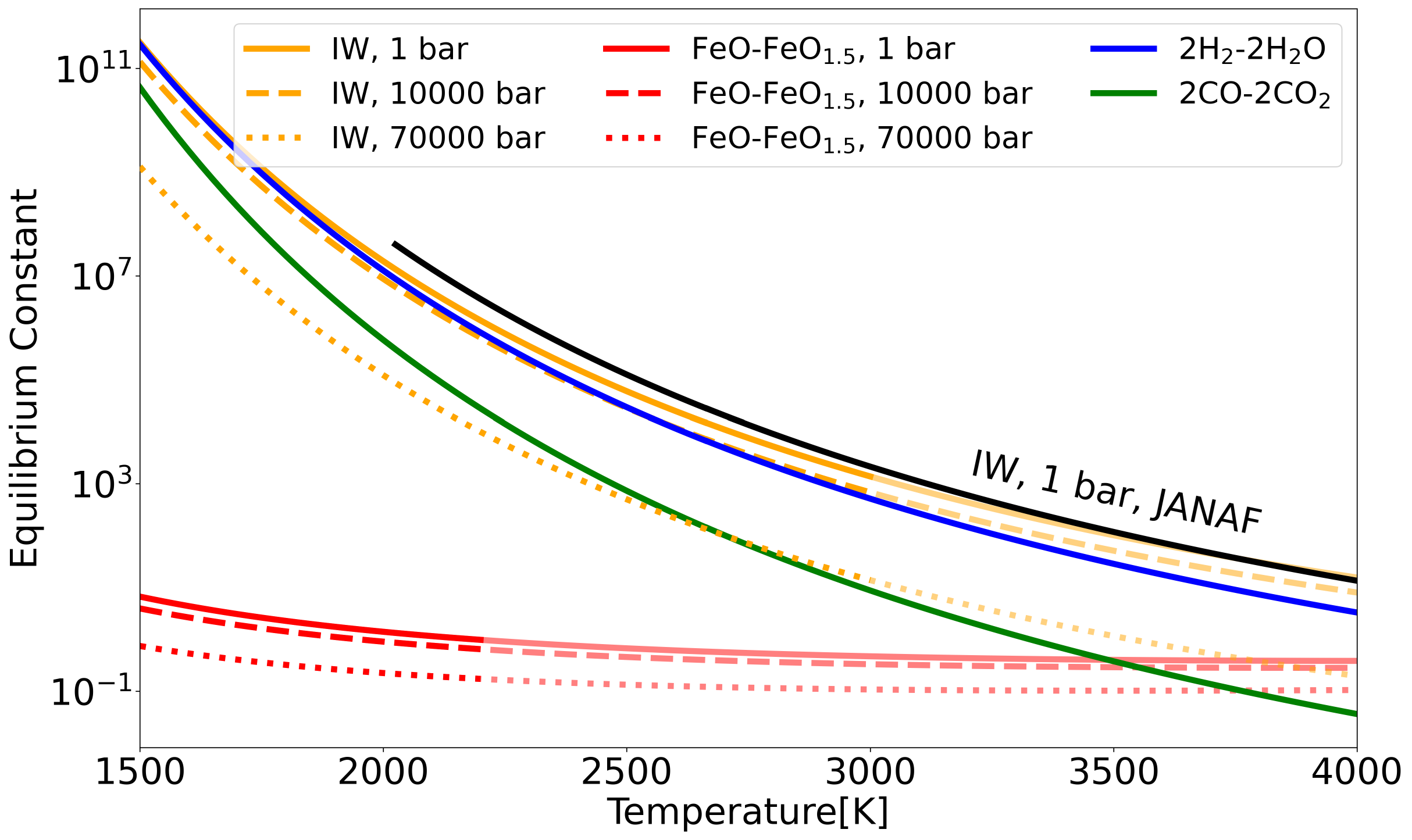}
\caption[Magma redox reactions]{The temperature-pressure dependencies of the equilibrium constants for significant reactions. The x-axis represents temperature, while the y-axis indicates the equilibrium constant. Different colored lines represent the equilibrium constants of individual reactions, as specified in the legend. The solid line corresponds to the default pressure of 1 bar, while dashed and dotted lines denote the IW and Ferric-Ferrous buffer equilibrium constants at pressures of 10000 bar and 70000 bar (the upper limit of the validation range). The lines with faded colors denote temperatures beyond the validated range, where extrapolation is applied. The black solid line represents the equilibrium constant of the IW buffer at 1 bar, calculated using the JANAF table.} 
\label{fig:Ks}
\end{figure}

Figure \ref{fig:Ks} illustrates the temperature and pressure dependencies of equilibrium constants for the IW buffer, Ferric-Ferrous equilibrium, \HH-\HHO, and \CO-\COO\ reactions. All reactions share similar temperature trends, becoming more reduced as temperature increases, although the sensitivities vary depending on the reactions. The IW buffer and Ferric-Ferrous equilibrium show pressure dependence when the pressure exceeds $\sim 10^4$ bars. It is worth mentioning that the \HH-\HHO\ and IW buffer exhibit similar temperature dependencies, suggesting a smaller temperature effect for the combined reaction (\HH + FeO $\leftrightarrows $ \HHO + Fe) compared to each reaction.

%%%%%%%%%%%%%%%%%%%%
%%%%%%%%%%%%%%%%%%%%

As shown in Figure \ref{fig:Ks}, the temperature validation ranges of the IW and Ferric-Ferrous buffers are much narrower compared to the \HH-\HHO\ and \CO-\COO\ reactions, extending only up to 3000~K and 2200~K. To discuss the results with higher surface temperature, we extrapolate the equilibrium constants of the two buffers as follows. We extrapolate the equilibrium constants up to 4000~K, based on the negligible difference (see Figure \ref{fig:Ks}) between the two IW equilibrium constants from JANAF and \citet{kowalski1995thermodynamic} and embracing arguments from \citet{2022GeCoA.328..221H} regarding the thermodynamic model of Ferric-Ferrous equilibrium. Beyond 4000~K, we employ boundary values for the equilibrium constants of all redox reactions. In Appendix \ref{ap:extrapolation}, we confirmed that the current extrapolation setting does not yield notable differences from the boundary-value extrapolation for all equilibrium constants over 2200~K (see Appendix \ref{ap:extrapolation} for details).

%%%%%%%%%%%%%%%%%%%%%%%%%%%%%%%%%%%%%%%%%%%%%%%%%%
\subsection{Dissolution of volatile species into magma}
\label{ss:model_dissolution}

Here we describe the solubility law for each species used in this paper.

We adopt the following law for the \HH\ solubility in basaltic silicate, based on the experiments conducted at temperatures of 1673 to 1723 K and pressures ranging from 7000 to 30000~bar (\citealt{2012E&PSL.345...38H}):
%%%%%%%%%%%%%%%%%%%%
\begin{equation}
X_{\rm H_{2}}=f\rm H_{2}[{\rm bar}]\, e^{-11.403-0.76P[{\rm GPa}]} \label{eq:solubility_H2_org}
\end{equation}
%%%%%%%%%%%%%%%%%%%%
where $X_{\rm H_{2}}$ is the \HH\ mole fraction, $f\rm H_{2}$ is the \HH\ fugacity (in bar), and $P$ is the atmospheric pressure (in GPa). 
This can be rewritten as:

%%%%%%%%%%%%%%%%%%%%
\begin{eqnarray}
X_{\rm H_{2}} &\sim& 0.11 \cdot s_{\rm H_{2}}(P) \left( \frac{P_{\rm H_{2}}}{1\,{\rm GPa}} \right)  \label{eq:solubility_H2} \\
s_{\rm H_{2}}(P) &\equiv & \phi_{\rm H_{2}}(0.47)^{P_{\rm } [{\rm GPa}]} 
\end{eqnarray}
%%%%%%%%%%%%%%%%%%%%

For the \HHO\ solubility to the basaltic magma, we refer to the equation from \citet{2016ApJ...829...63S}. They applied a curve fit to the \HHO\ solubility data for various silicate melts from \citet{1997CoMP..126..237P} which has a temperature range of 900 to 1670 K and a pressure range between 100 and 10000~bar:

%%%%%%%%%%%%%%%%%%%%
\begin{eqnarray}
\begin{split}
% X_{H2O}=\sqrt{1.05*10^{-4}fH_{2}O},  
w_{\rm H_{2}O} &=&  3.44 \cdot 10^{-8} (P_{\rm H_{2}O}[{\rm Pa}])^{0.74 } \label{eq:solubility_H2O_org}
\end{split}
\end{eqnarray}
%%%%%%%%%%%%%%%%%%%%
where $w_{\rm H_{2}O}$ is the weight fraction of \HHO, and $P_{\rm H_{2}O}$ is the \HHO\ partial pressure. 
This may be rewritten as:
%%%%%%%%%%%%%%%%%%%%
\begin{eqnarray}
% X_{H2O}=\sqrt{1.05*10^{-4}fH_{2}O},  
X_{\rm H_{2}O} &=&  0.16 \cdot s_{\rm H_{2}O}\left( \frac{P_{\rm H_{2}O}}{1\,{\rm GPa}} \right)^{0.74 }  \label{eq:solubility_H2O_org} \\
s_{\rm H_{2}O} &\equiv & \left( \frac{\mu_{\rm m}}{\mu_{\rm H_{2}O}} \right) \label{eq:solubility_H2O}
\end{eqnarray}
%%%%%%%%%%%%%%%%%%%%
where $\mu_{\rm m}$ is the molar mass of magma. 
Note that the solubility of \HHO\ does not linearly depend on the ${P_{\rm H_2O}}$. It implies the \HHO\ dissolution into silicate melt is not completely physical dissolution, and the chemical reaction between the magma and \HHO\ fluid changes the solubility law. More specifically, \HHO\ has two dissolution mechanisms in silicate \cite[e.g.,][]{1996JGR...10114909K, 1999JVGR...88..201Z, 2005JVGR..143..219L, 2017CoMP..172...85M, 2023FrEaS..1159412S}:

%%%%%%%%%%
\begin{itemize}
    \item $\textrm{H}_{2}\textrm{O}_{\textrm{fluid}}\ \leftrightarrows\ \textrm{H}_{2}\textrm{O}_{\textrm{melt}}$
    \item $\textrm{H}_{2}\textrm{O}_{\textrm{melt}}\ +\ \textrm{O}^{2-}_{\textrm{melt}} \leftrightarrows  2{\textrm{OH}^{-}}_{\textrm{melt}}$
\end{itemize}
%%%%%%%%%%

The former reaction is the physical dissolution of \HHO\ into the melt, which follows Henry's law, while the latter reaction establishes a proportionality between the dissolved \HHO\ and the square root of partial pressure of \HHO\ ($\propto P_{\rm H_{2}O}^{0.5}$). Therefore, the solubility of \HHO\ has an exponent between 0.5 and 1, with the exact value depending on the dissolved amount of $\rm OH^{-}$ in the magma. A high concentration of hydroxide ions in magma makes the physical dissolution of \HHO\ more significant. This transition is understood to occur at a boundary of a few weight percentages of dissolved \HHO\ \cite[e.g.,][]{1982GeCoA..46.2609S, 1997CoMP..126..237P, 2017CoMP..172...85M, 2023FrEaS..1159412S}.

The solubility law for \COO\ is adopted from the fitting function given by \citet{1991GeCoA..55.1587P}, which has a pressure range of 1000 to 15000~bar and a temperature range of 1473 to 1873 K. The original equation is :

%%%%%%%%%%%%%%%%%%%%
\begin{eqnarray}
\begin{split}
w_{\rm CO_{2}} [\%] = 0.00119 + 4.81438 \cdot 10^{-5} \cdot P_{\rm CO_{2}}[\rm bar]\\
+ 5.019505 \cdot 10^{-9} \cdot P^{2}_{\rm CO_{2}}\\
- 2.587138 \cdot 10^{-13} \cdot P^{3}_{\rm CO_{2}}\\
+ 5.96362 \cdot 10^{-18} \cdot P^{4}_{\rm CO_{2}}\\
- 5.67816 \cdot 10^{-23} \cdot P^{5}_{\rm CO_{2}}
\end{split}
\end{eqnarray}
%%%%%%%%%%%%%%%%%%%%
where $w_{\rm CO_{2}}$ is the weight fraction of \COO. However, the above fitting function may not be a good approximation at small $P_{\rm CO_{2}}$ because the mass fraction of \COO\ in silicate does not converge to zero, even at zero bar. Therefore, we use a re-fitted function that employs only a first-order term directly proportional to the partial pressure of \COO. Note that such a linear dependence on $P_{\rm CO_{2}}$ is expected for the following dissolution mechanism in an idealized situation ($\rm CO_{2,fluid}\ +\ \rm{O}^{2-}_{melt} \leftrightarrows {CO}_{3,melt}^{2-}$; \citealt{1991GeCoA..55.1587P}):

%%%%%%%%%%%%%%%%%%%%
\begin{eqnarray}
w_{\rm CO_{2}} = 0.0054P_{\rm CO_{2}}[{\rm GPa]} \label{eq:solubility_CO2_org} 
\end{eqnarray}
%%%%%%%%%%%%%%%%%%%%
where $w_{\rm CO_{2}}$ is the weight fraction of \COO. This equation shows a below 5\% difference from the original equation within the pressure range of 200 to 1700~bar. We rewrite this equation to:
%%%%%%%%%%%%%%%%%%%%
\begin{equation}
X_{\rm CO_{2}} = 0.0054 \left( \frac{\mu_{\rm m}}{\mu_{\rm CO_{2}}} \right) \left( \frac{P_{\rm CO_{2}}}{1\,{\rm GPa}} \right).\label{eq:solubility_CO2}
\end{equation}
%%%%%%%%%%%%%%%%%%%%

The \CO\ dissolution to MORB satisfies the following equation at temperatures between 1523 and 1873~K and pressures between 2000 and 30000~bar (\citealt{2019GeCoA.259..129Y}): 
%%%%%%%%%%%%%%%%%%%%
\begin{equation}
w_{\rm CO} [\mbox{\%}]=10^{-4.83+0.8 \log f \rm CO [{\rm bar}]}, \label{eq:solubility_CO_org}
\end{equation}
%%%%%%%%%%%%%%%%%%%%
or
%%%%%%%%%%%%%%%%%%%%
\begin{equation}
X_{\rm CO} = 0.0002 \cdot (\phi_{\rm CO})^{0.8} \cdot \left( \frac{\mu_{\rm m}}{\mu_{\rm CO}} \right)\left( \frac{P_{\rm CO}}{1\,{\rm GPa}} \right)^{0.8}. \label{eq:solubility_CO}
\end{equation}
%%%%%%%%%%%%%%%%%%%%

The \CHHHH\ solubility law refers to the solubility to haplo-basalt given in \citet{2013GeCoA.114...52A} with the experimental data with pressures within 7000 to 30000~bar and temperatures between 1673 and 1723~K. We omit the temperature dependence of \CHHHH\ solubility due to their limited temperature range (e.g., \citealt{2012E&PSL.345...38H}). The derived solubility law is as follows:

%%%%%%%%%%%%%%%%%%%%
\begin{equation}
X_{\rm CH_{4}}=f\rm CH_{4}[{\rm GPa}] \, e^{-7.63-1.9P[{\rm GPa}]} \label{eq:solubility_CH4_org}, 
\end{equation}
%%%%%%%%%%%%%%%%%%%%
or
%%%%%%%%%%%%%%%%%%%%
\begin{eqnarray}
X_{\rm CH_{4}} &=& 0.0005 \cdot s_{\rm CH_{4}} \left( \frac{P_{\rm CH_{4}}}{1\,{\rm GPa}} \right) \label{eq:solubility_CH4} \\
s_{\rm CH_{4}} &=& \phi_{\rm CH_{4}}(0.15)^{P[{\rm GPa}]} 
\end{eqnarray}
%%%%%%%%%%%%%%%%%%%%

At this point, it is worth estimating the impact of dissolution into magma on the atmospheric composition by approximately calculating the fraction of dissolved species in the magma layer relative to those in the atmosphere at the BOA. For this, we use the following approximate relations for the total number of molecules in the atmosphere, $N_{\rm atm}$:
%%%%%%%%%%%%%%%%%%%%
\begin{eqnarray}
N_{\rm atm} &\sim & \frac{4 \pi R_{\rm core}^2 P_{\rm BOA}}{\mu_{\rm atm} g }\\
&\sim &  \frac{4 \pi R_{\rm core}^2 P_{\rm BOA}}{\mu_{\rm atm} } \frac{R_{\rm core}^2}{GM_{\rm core}} \\
&\sim & \frac{4 \pi R_{\rm core}^4 P_{\rm BOA}}{GM_{\rm p} \mu_{\rm atm} } 
\end{eqnarray}
%%%%%%%%%%%%%%%%%%%%
where $P_{\rm BOA}$ is the pressure at BOA and $\mu _{\rm atm}$ is the mean molecular weight of the atmosphere. 
Here, we ignore the dependence of the gravity $g$ on the elevation above the magma-atmosphere interface. Meanwhile, the total number of molecules in magma, $N_{\rm m}$, can be written as
%%%%%%%%%%%%%%%%%%%%
\begin{eqnarray}
N_{\rm m} = \frac{ M_{\rm core} \zeta  }{\mu_{\rm m}} \sim \frac{ M_{\rm p} \zeta  }{\mu_{\rm m}},
\end{eqnarray}
%%%%%%%%%%%%%%%%%%%%
where $\zeta$ is the mass fraction of the silicate layer that is molten and reactive. Denoting the mixing ratio of species $S$ in the atmosphere is indicated by $x_S$, the ratio of the species $S$ dissolved in magma to those in the atmosphere, $r_S$, can be expressed as follows:
%%%%%%%%%%%%%%%%%%%%
\begin{eqnarray}
r_S &=& \frac{ N_{\rm m} X_S }{N_{\rm atm} x_S} \\
&=& \frac{ G M_{\rm p}^2 \zeta  }{4\pi R_{\rm core}^4} \frac{\mu_{\rm atm}}{\mu_{\rm m}} \frac{X_S}{x_S P_{\rm BOA} }  \\
&\sim& 42 \, \left( \frac{ M_{\rm p} }{6M_{\oplus }} \right)^{2} \left( \frac{ R_{\rm p} }{1.6R_{\oplus }} \right)^{-4} \left( \frac{\zeta }{0.66} \right) \left( \frac{\mu_{\rm atm}}{0.1 \mu_{\rm m}} \right) \frac{X_S}{P_{{\rm BOA},S}[{\rm GPa}]}, \label{eq:atm_diss_ratio}
\end{eqnarray}
%%%%%%%%%%%%%%%%%%%%
where $P_{\rm BOA,S}$ is the partial pressure of the species $S$ at BOA. Substitution of Equations (\ref{eq:solubility_CO}) and (\ref{eq:solubility_CH4}) indicates that $r_{\rm CH_{4}}$ and $r_{\rm CO}$ range from about 0.01 to 0.1 if the atmospheric pressure is lower than a few GPa. This implies that the reduced C-bearing species (\CO\ and \CHHHH) will remain in the atmosphere in most cases (neglecting uptake by the iron core).

The solubility of \COO\ could marginally affect the result:
%%%%%%%%%%%%%%%%%%%%
\begin{eqnarray}
r_{\rm CO2} &=& 0.23\, \left( \frac{ M_{\rm p} }{6M_{\oplus }} \right)^{2} \left( \frac{ R_{\rm p} }{1.6R_{\oplus }} \right)^{-4} \left( \frac{\zeta }{0.66} \right) \left( \frac{\mu_{\rm atm}}{0.1\mu_{\rm CO_{2}}} \right). 
\label{eq:rCO2}
\end{eqnarray} 
%%%%%%%%%%%%%%%%%%%%

Thus, only in the case where the atmosphere is highly oxidized ($\mu_{\rm atm} \gtrsim 0.1 \mu_{\rm CO_{2}}$), and the substantial fraction of silicate is almost fully molten and well-mixed, i.e., $\zeta \sim 0.66$, atmospheric CO$_2$ is somewhat affected by the dissolution.  

We do not set an upper limit on the dissolved mole fraction of volatiles in magma. Within our parameter range, the maximum mole fraction of volatiles in the magma reaches about 29\%, corresponding to a mass fraction of about 4.0\%. Given the small mass fraction of the volatile species in magma, we neglect the density decrease and compositional changes in magma due to dissolved volatiles, as well as their effects on chemical reactions and the mass-radius relation of the core.

\subsection{Numerical procedure}
\label{ss:numerical_procedure}

We calculate the equilibrium atmospheric compositions of the atmosphere-magma system in terms of five species (\HH, \HHO, \CHHHH, \CO, \COO). The elemental composition of the total accreted volatile species matches the solar system composition from \citet{2003ApJ...591.1220L}; C/H$= 2.9 \times  10^{-4}$ and C/O$= 0.5$. This implicitly assumes that nebula gas accretion took place interior to the snowline of \HHO\ and C-bearing species, which would be consistent with the assumption of the rocky interior \cite[e.g.,][]{2021JGRE..12606639B}. The five species both in the atmosphere and the magma ($2\times 5=10$ variables) and the amount of O bounded to Fe in the rock (1 variable) can be determined by the three redox equations, the five dissolution laws, and the total amount of H, O, and C in the atmosphere-magma system (three constraints); see also \citet{2022PSJ.....3..127S}. 

The full set of equations is inherently non-linear. 
The non-linearity would disappear when we assume that (1) the atmospheric amount of \CO, \COO, and \CHHHH\ is negligible, and (2) the non-linear solubility law of \CO\ is negligible, both of which are valid in the case studied here where the C/H in the accreted volatile is small. However, for future applications to general cases, we solve the equations using a scipy Nelder-Mead root-finding algorithm (\citealt{2020SciPy-NMeth}) in the following way.

\begin{enumerate}

\item We begin with estimating the radius of the planetary core $R_{\rm core}$ and the number of molecules of the magma layer $N_{\rm m}$, using the mass-radius relationship of bare rock by \citet{2007ApJ...659.1661F} with the given mass fraction of the silicate layer (default: 0.66) and the externally given planetary mass, $M_{\rm p}$. Note that we ignore the mass of the atmosphere since the mass fraction of the atmosphere for sub-Neptunes considered here is at most 0.46\%. 
The difference between the planetary radius $R_{\rm p}$ and $R_{\rm core}$ gives the thickness of the atmosphere, $\Delta R$. 

\item We then prescribe a value for the atmospheric \HHO\ fraction $x_{\rm H_{2}O}$. The much smaller amount of C compared to H ($\sim 10^{-4}$) and the fixed atmospheric $\rm \frac{He}{H}$ ratio allow us to derive the atmospheric \HH\ fraction $x_{\rm H_{2}}$ and the He fraction $x_{\rm He}$ from $x_{\rm H_{2}O}$. The temperature and pressure at the BOA are either specified directly (Section \ref{ss:atmospheric composition}) or calculated using the adiabatic lapse rate for the given atmospheric composition and planetary parameters (Section \ref{ss:atmospheric composition_magmaparamdepen}).

\item The oxygen fugacity ($f \rm O_2$) and rock composition (the number of pure Fe in magma, [\FeO], [\FeFeOOO]) are determined based on the redox equilibrium given the temperature and pressure at BOA as well as the $\frac{x_{\rm H_{2}O}}{x_{\rm H_{2}}}$. 

\item From the composition of the accreted volatiles (with the abundance ratio of H, O, and C being the solar values), surface temperature ($T_{\rm BOA}$), surface pressure ($P_{\rm BOA}$), $f\textrm{O}_{\rm2}$, atmospheric He/H ratio, $N_{\rm m}$, and rock composition, we determine the amounts of \HH, \HHO, \CO, \COO, \He, and \CHHHH\ in the atmosphere and magma under the constraints on the O/H and C/H of the accreted gas. We employ a root-finding algorithm on the atmospheric C/H ratio due to the non-linearity. If the atmospheric C/H ratio is given, atmospheric amounts of C-bearing species ($x_{\rm CO}$, $x_{\rm CO_{2}}$, and $x_{\rm CH_{4}}$) can be derived from $x_{\rm H_{2}}$, $x_{\rm H_{2}O}$, and surface conditions ($T_{\rm BOA}$, $P_{\rm BOA}$, and $f\textrm{O}_{\rm2}$). Then, solubility laws and $N_{\rm m}$ give the dissolved amounts of all volatiles, and planetary amounts of all volatiles naturally emerge. The root-finding algorithm finds the atmospheric C/H ratio that makes the planetary volatile composition consistent with the assumed nebula composition: C/H$= 2.9 \times 10^{-4}$. To be specific, the difference between the calculated planetary C/H and the assumed nebula C/H is minimized until it becomes smaller than $10^{-5}$.

\item The equilibrium calculation gives the total amount of O ($N_{\rm O}$). The difference between $N_{\rm O}$ and the amount of O from accreted gas is interpreted as the amount of $N_{\rm O(Fe)}$ before the chemical reaction, which is also related to the redox state of magma before the gas accretion.

\end{enumerate}

By repeating this process with varying $x_{\rm H2O}$, we obtain the relation among $x_{\rm H2O}$, the mixing ratios of C-bearing species, and the rock composition (both before and after the gas accretion). 

\subsection{Consideration of magma solidification}
\label{ss:magma_solidification}

When considering magma solidification, we assume a scenario where the atmosphere first reacts with the initially hot, fully molten homogeneous silicate core, and the chemical equilibrium changes as the magma solidifies from the bottom over time. In this study, we simplify the situation by considering only two discrete stages rather than the continuous solidification and the change in equilibrium state: (1) the initial fully molten magma reacting with the atmosphere and (2) the completion of solidification (i.e., the desired solidified fraction has been reached). We also assume that there is no exchange of material with the outside of the magma-atmosphere system and that the surface temperature remains constant. The concrete calculation method is as follows:

\begin{enumerate}

\item We calculate the chemical equilibrium state of a planet with fully molten homogeneous magma by assuming specific values for the atmospheric \HHO\ fraction $x_{\rm H_{2}O}$, $T_{\rm BOA}$, and $P_{\rm BOA}$. From this result, we determine the planetary amounts of H, C, He, and O (Section \ref{ss:numerical_procedure}). 

\item We find the new equilibrium state under the updated molten fraction. As mentioned in Section \ref{ss:model_structure}, we assume that no volatiles can exist in solidified magma, so the amounts of H, C, and He in the magma and atmosphere remain the same as in the initial state. However, some of the O is excluded from the reaction in the form of \FeO\ and \FeFeOOO\ contained in the solidified magma. The excluded amount can be calculated from the iron speciation of the magma obtained in step 1. Again, a root-finding algorithm is applied to determine the $P_{\rm BOA}$, atmospheric C/H ratio, atmospheric He/H ratio, and $x_{\rm H_{2}O}$ that satisfy the given amounts of H, C, He, and O.  

\end{enumerate}

\subsection{Parameter range}
\label{ss:Paramrange}

In order for some of our assumptions to be valid, we consider the results only when the following criteria are met:
\begin{itemize}
\item $T_{\rm BOA} > 1800$~K: This ensures that the silicate layer is at least partially molten.
\item $P_{\rm BOA} < 30000$~bar: This range is determined based on the pressure range of the solubility experimental study of the most dominant gas species in our results, \HH. With our fiducial set of planetary parameters ($M_{\rm p}=6M_{\oplus }$, $T_{\rm eq}=750~\rm K$) and assuming an \HH-rich atmosphere ($x_{\rm H_{2}O}=0.01$), 30000~bar corresponds to the atmospheric mass fraction of $0.5\%$ and the atmospheric thickness of $\Delta R \sim 0.8R_{\oplus}$, corresponding to $R_{\rm p}=2.4R_{\oplus}$. This provides an idea of the upper limit of the planetary radius discussed in this paper. Due to the lack of available data, we do not explicitly discuss larger planets such as K2-18~b, although we expect qualitatively similar trends.  
\end{itemize}

\section{Result}
\label{s:result}
\subsection{Dependence on surface conditions and magma redox state}
\label{ss:atmospheric composition}

This section presents the relation between atmospheric composition and magma redox state before the reaction (as represented by the iron speciation of the magma), with a specific focus on the influence of two fundamental parameters in determining the equilibrium state: surface pressure and temperature. Here, the rocky layer is assumed to be fully molten.

%%%%%%%%%%%%%%%%%%%%
\begin{figure*}
\centering
\includegraphics[width=18cm]{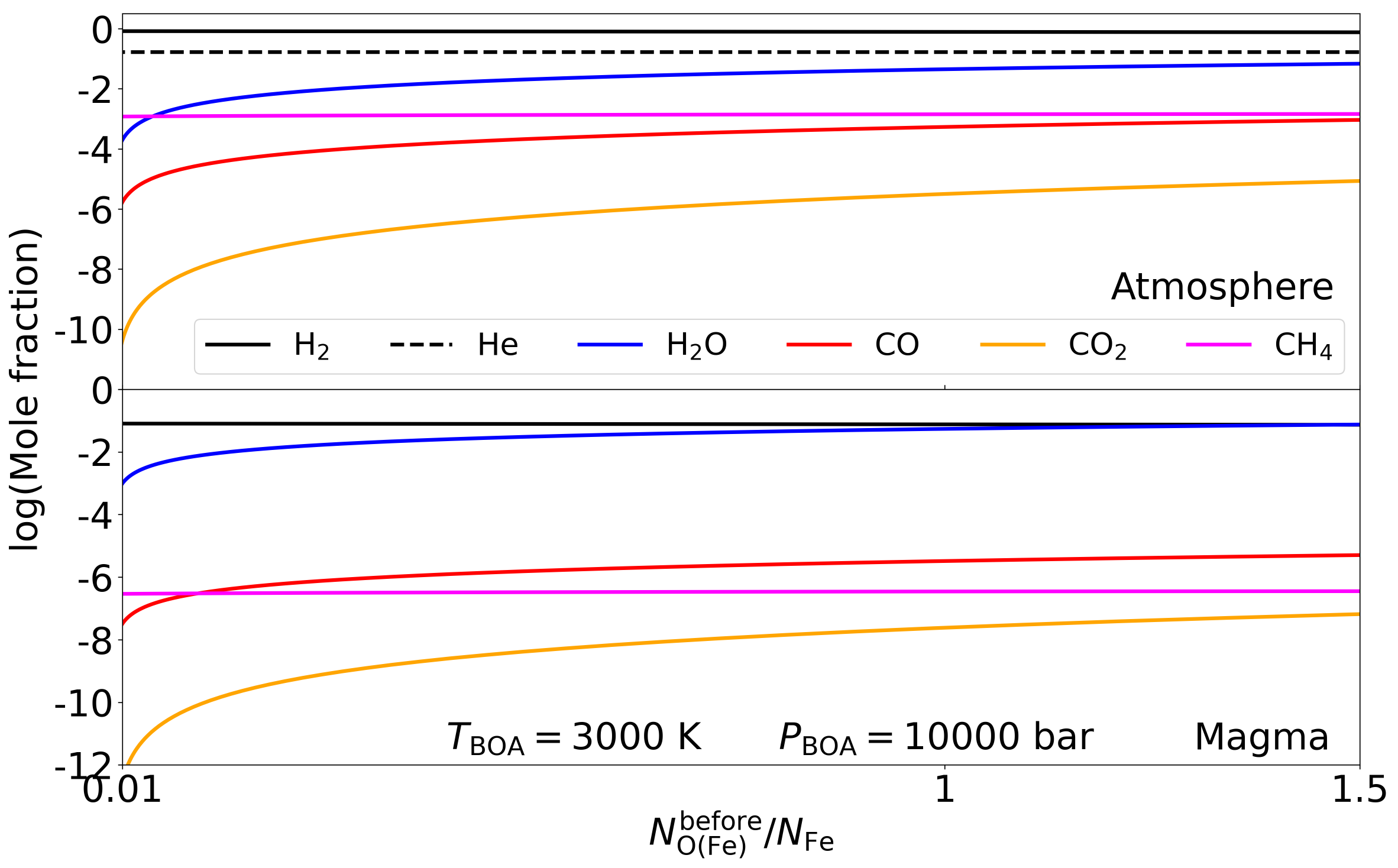}
\caption{The relation between the mole fraction of each species and $N_{\rm O(Fe)}^{\rm before}$/$N_{\rm Fe}$. The variable for the x-axis, $N_{\rm O(Fe)}^{\rm before}$/$N_{\rm Fe}$, represents the redox state of the silicate before the reaction. The upper panel corresponds to the atmospheric composition, while the lower panel denotes the species present in the magma. The surface temperature and pressure are fixed at 3000~K and 10000~bar, respectively.}
\label{fig:atmos_magma_species_Oini}
\end{figure*}
%%%%%%%%%%%%%%%%%%%%

Before discussing elemental ratios, we first introduce how volatile speciation depends on the magma redox state. Figure \ref{fig:atmos_magma_species_Oini} shows the volatile speciation in the planetary model as a function of the magma redox state before reactions, $N_{\rm O(Fe)}^{\rm before}$/$N_{\rm Fe}$. The upper and lower panels represent the atmosphere and magma, respectively. Surface pressure ($P_{\rm BOA}$) and temperature ($T_{\rm BOA}$) are fixed at 10000 bar and 3000 K. It is clear that as the magma becomes more oxidized, the fractions of \HHO, \CO, and \COO\ increase relative to \HH\ and \CHHHH. Additionally, the larger solubility of H-bearing species, particularly \HHO, results in their relative mole fractions in magma being substantially larger than those of C-bearing species.

%%%%%%%%%%%%%%%%%%%%
\begin{figure*}
\centering
\includegraphics[width=18cm]{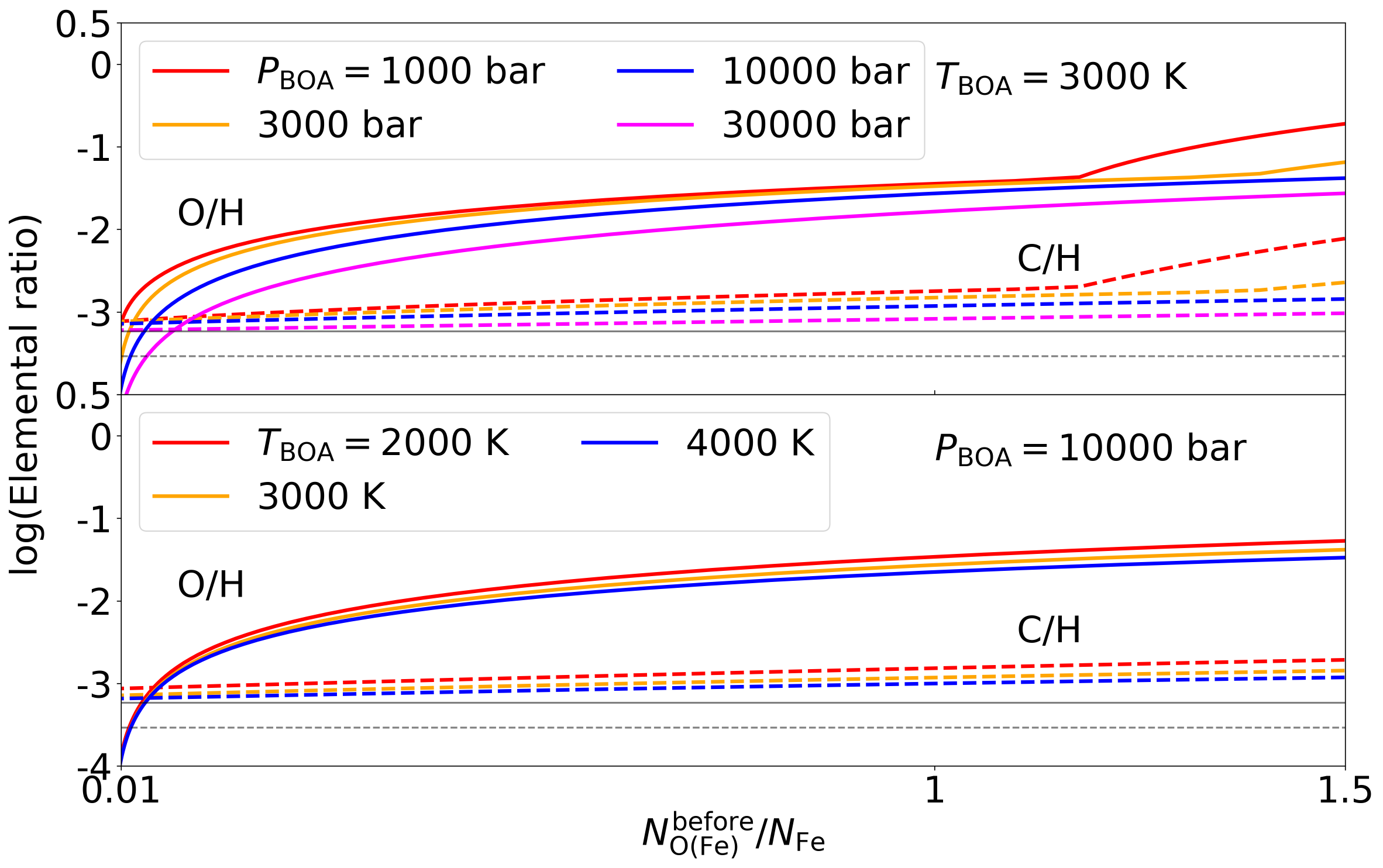}
\caption{The relation between the atmospheric H-C-O ratio and $N_{\rm O(Fe)}^{\rm before}$/$N_{\rm Fe}$, with its dependency on surface pressure and temperature. The solid line corresponds to the O/H ratio, while the dashed line represents the C/H ratio. We note that $\rm log(O/H)$ and $\rm log(C/H)$ of accreted volatiles are assumed to be about $-3.23$ and $-3.53$. (Upper) We show four surface pressure cases: 1000, 3000, 10000, and 30000 bar. The surface temperature is fixed at 3000 K. (Lower) Three surface temperature cases are shown: 2000, 3000, and 4000 K. The surface pressure is fixed at 10000 bar.}
\label{fig:atmos_Oini_linear_surfacedepen}
\end{figure*}
%%%%%%%%%%%%%%%%%%%%

Figure \ref{fig:atmos_Oini_linear_surfacedepen} shows the atmospheric O/H and C/H as a function of $N_{\rm O(Fe)}^{\rm before}$/$N_{\rm Fe}$. The upper and lower panels represent the dependence on surface pressure ($P_{\rm BOA}$) and temperature ($T_{\rm BOA}$), respectively. 

The atmospheric O/H ratio is primarily governed by the redox state of the silicate before the reaction, with its dependence on BOA pressure becoming non-negligible as pressure increases. When the pressure exceeds approximately 30000 bar, the atmospheric O/H ratio decreases to about half of what it is at 1000 bar under the condition where $N_{\rm O(Fe)}^{\rm before}$/$N_{\rm Fe} \sim 1$. Overall, the atmospheric \HHO\ fraction ($x_{\rm H_{2}O} \sim 2\frac{\rm O}{\rm H}$) ranges from a few percentages to a few tens of percentages (for thin atmospheres with $P_{\rm BOA} \lesssim \rm 3000~bar$) unless the magma before the reaction is highly reduced, where $N_{\rm O(Fe)}^{\rm before}$/$N_{\rm Fe} \lesssim 0.1$. On the other hand, the atmospheric C/H ratio appears relatively insensitive to $N_{\rm O(Fe)}^{\rm before}$/$N_{\rm Fe}$ and surface pressure. The atmospheric C/H ratio maintains the nebula gas composition except in the lowest-pressure ($< 1000$~bar) and most oxidized atmospheres with an \HHO\ mixing ratio exceeding a few percent, leading to an enrichment of C/H several dozen times. The impact of temperature on both O/H and C/H ratios appears to be relatively limited, at least for temperatures below 4000~K. 

This can be understood from an approximate analytic expression for O/H and C/H, described below. We first focus on the O/H ratio.  
We begin by writing a simple mass conservation of O:
%%%%%%%%%%%%%%%%%%%%
\begin{equation}
N_{\rm O(Fe)}^{\rm before} = N_{\rm O(Fe)}^{\rm after} + N_{\rm atm} x_{\rm H_{2}O} + N_{\rm m} X_{\rm H_{2}O}, \label{eq:O_conservation}
\end{equation}
%%%%%%%%%%%%%%%%%%%%
where $N_{\rm atm}$ and $N_{\rm m}$ represent the total number of molecules in the atmosphere and magma, respectively. However, due to the large solubility of \HHO\ in magma, the second term on the right-hand is substantially smaller than the third term if $\frac{\zeta }{0.66} \gtrsim 0.5$ under our fiducial planetary parameter combination (See Equation \ref{eq:H2Osol}). Therefore, for simplicity, the second term is omitted below.

Using Equation (\ref{eq:solubility_H2O_org}) and assuming that the Iron-W\"{u}stite buffer dominates the reactions after the reaction ([FeO]$^{\rm after}$ $\gg $ [Fe$_2$O$_3$]$^{\rm after}$),

%%%%%%%%%%%%%%%%%%%%
\begin{eqnarray}
N_{\rm O(Fe)}^{\rm before} &=& N_{\rm m} \left\{ {\rm [FeO]}^{\rm after} 
%+ N_{\rm a} x_{\rm H2O} \notag \\
+ 0.16 \left( \frac{\mu_{\rm m}}{\mu_{\rm H_{2}O}} \right) \left( \frac{P_{\rm H_{2}O}}{\rm 10000~bar} \right)^{0.74} \right\}
\label{eq:conservation_O}
\end{eqnarray}
%%%%%%%%%%%%%%%%%%%%
where $\frac{\mu_{\rm m}}{\mu_{\rm H_{2}O}}\sim 3.4$ from the assumed magma composition.

Denoting the equilibrium constant of the following reaction: \HH + FeO $\leftrightarrows $ \HHO + Fe, we may write 
%%%%%%%%%%%%%%%%%%%%
\begin{eqnarray}
K &=& \frac{f\rm H_{2}O{\rm [Fe]}}{\gamma_{\rm FeO}f\rm H_{2}{\rm [FeO]}} \\
&=&  \frac{\phi_{\rm H_{2}O}x_{\rm H_{2}O}}{\gamma_{\rm FeO}\phi_{\rm H_{2}}x_{\rm H_{2}}{\rm [FeO]}}.
\end{eqnarray}
%%%%%%%%%%%%%%%%%%%%

\begin{figure}
\centering
\includegraphics[width=8.5cm]{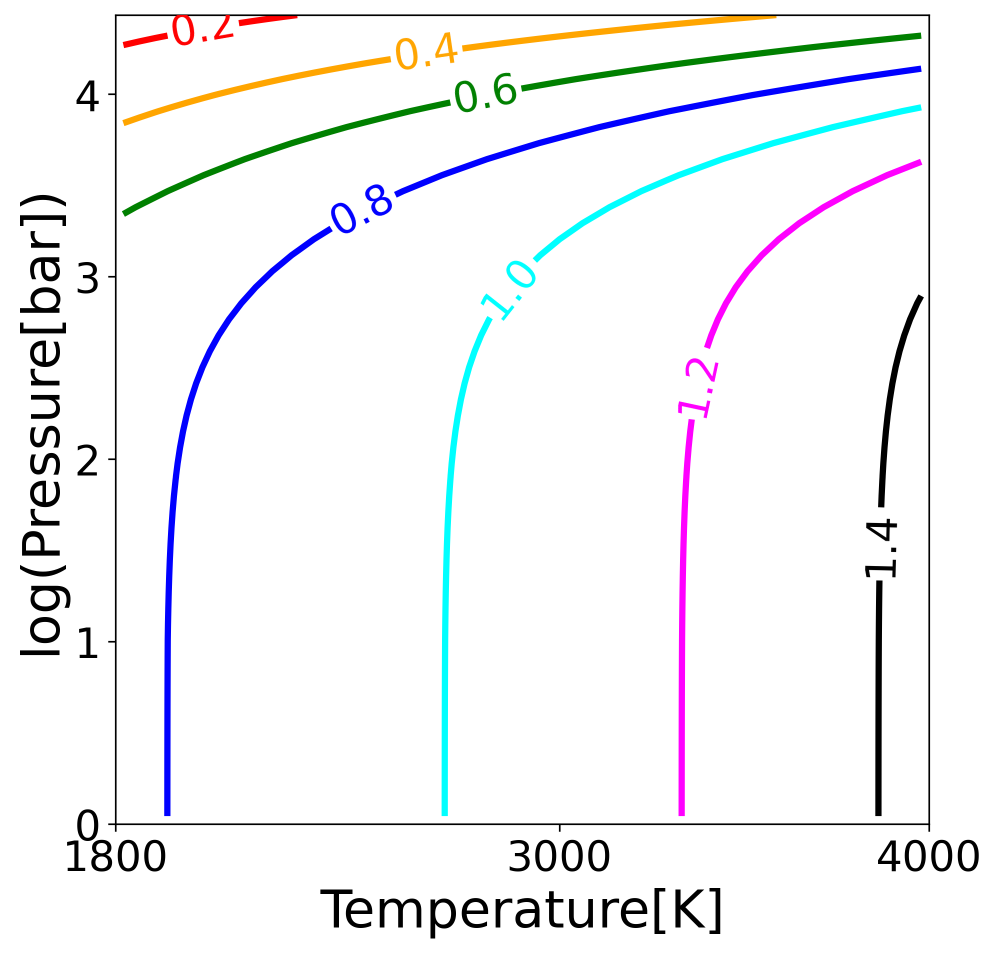}
\caption{The temperature- and pressure-dependence of the $K'=\frac{\phi_{\rm H_{2}O}}{\gamma_{\rm FeO}\phi_{\rm H_{2}}K}$.} 
\label{fig:K}
\end{figure}

Therefore, Equation (\ref{eq:conservation_O}) becomes

%%%%%%%%%%%%%%%%%%%%
\begin{eqnarray}
N_{\rm O(Fe)}^{\rm before} &=& N_{\rm m} \left( a_1 x_{\rm H_{2}O} + a_2  x_{\rm H_{2}O}^{0.74} \right) \label{eq:conservation_O_modified}\\
\frac{N_{\rm O(Fe)}^{\rm before}}{N_{\rm m}} = \frac{N_{\rm O(Fe)}^{\rm before}}{N_{\rm Fe}} f_{\rm Fe} &=& a_1 x_{\rm H_{2}O} + a_2  x_{\rm H_{2}O}^{0.74} \label{eq:conservation_O_modified} \\
a_1 &\equiv & 1.2\frac{\phi_{\rm H_{2}O}}{\gamma_{\rm FeO}\phi_{\rm H_{2}}K} \\
a_2 &=& 0.54 \left( \frac{P_{\rm BOA}}{\rm 10000~bar} \right)^{0.74}, 
\end{eqnarray}
%%%%%%%%%%%%%%%%%%%%
where $f_{\rm Fe}$ is the mole fraction of Fe in magma. We note that $\frac{1}{x_{\rm H_{2}}} \sim 1.2$ under the condition that \HH\ and He dominate the atmospheric composition, and the He/H ratio is 0.1. The first term makes the relation $N_{\rm O(Fe)}^{\rm before} \propto x_{\rm H_{2}O}$. The pressure and temperature dependence of the part of the proportionality constant, $K'=\frac{\phi_{\rm H_{2}O}}{\gamma_{\rm FeO}\phi_{\rm H_{2}}K}$, is shown in Figure \ref{fig:K}. While $K'$ increases with increasing temperature and decreases with increasing pressure, $K'$ does not vary significantly. On the other hand, the second term leads to $N_{\rm O(Fe)}^{\rm before} \propto x_{\rm H _{2}O}^{0.74}$. The proportionality constant for this term is influenced solely by surface pressure, with higher pressure resulting in a larger constant. These assessments suggest that the second term becomes dominant for massive ($P_{\rm BOA}$$\gtrsim $ $10000$~bar) and reduced atmospheres.

%%%%%%%%%%%%%%%%%%%%
%%%%%%%%%%%%%%%%%%%%

Although these two terms have opposite trends in response to pressure change (i.e., an increase in pressure leads to a decrease in the first term and an increase in the second term), their sum simply increases as the pressure increases for a given $x_{\rm H_{2}O}$, as indicated by the upper panel of Figure \ref{fig:atmos_Oini_linear_surfacedepen}. 
This is because the pressure dependence of $K'$ is minimal under low pressure, where the first term dominates. This result is consistent with the intuitive understanding that as the atmospheric amount increases, $x_{\rm H_{2}O}$ decreases due to the limited oxidizing power of magma.

The temperature dependence of $K'$ results in the trend that the same $N_{\rm O(Fe)}$ corresponds to a more reduced atmosphere under higher temperatures. However, due to the limited variation in the value of $K'$ mentioned above, the overall dependence is minor, as depicted in the bottom panel of Figure \ref{fig:atmos_Oini_linear_surfacedepen}. 

Figure \ref{fig:atmos_Oini_linear_surfacedepen} shows that the slope of the O/H ratio--$N_{\rm O(Fe)}^{\rm before}$ relationship changes towards the most oxidized end at low pressure (1000 and 3000 bar). This abrupt change corresponds to the depletion of pure iron in magma, where the reaction transitions from $\textrm{Fe} \longleftrightarrow \textrm{FeO} \longleftrightarrow \textrm{Fe}_{2}\textrm{O}_{3}$ to $\textrm{FeO} \longleftrightarrow \textrm{Fe}_{2}\textrm{O}_{3}$ after the depletion of pure iron. This corresponds to the change in the slope of the relation between $f\rm O_{2}$ and $N_{\rm O(Fe)}$, shown in Figure \ref{fig:FeFeOFe2O3relation}. When the atmospheric pressure is higher than $\sim $3000~bar, the reducing power of the atmosphere is enough to produce pure iron, even if the rock before the reaction is fully oxidized. Therefore, a sudden change in the slope of the O/H ratio--$N_{\rm O(Fe)}^{\rm before}$ relationship is not observed.

Next, we will discuss the dependence of C/H on planetary parameters. As expected, the atmospheric C/H ratio is always larger than the input (total) C/H value ($10^{-3.53}$) due to the dissolution of H-bearing species into magma. Combining Equation (\ref{eq:atm_diss_ratio}) and the solubility laws, the ratio of the dissolved \HH\ and \HHO\ to the atmospheric counterparts are:
\begin{eqnarray}
r_{\rm H_{2}} &=& 4.6 \, \left( \frac{ M_{\rm p} }{6M_{\oplus }} \right)^{2}\left( \frac{ R_{\rm core} }{1.6R_{\oplus }} \right)^{-4} \left( \frac{\zeta }{0.66} \right)\left( \frac{\mu_{\rm atm}}{0.1 \mu_{\rm m}} \right)\phi_{\rm H_{2}}(0.47)^{P_{\rm BOA}[{\rm GPa}]} 
\label{eq:H2sol}\\
r_{\rm H_{2}O} &=& 22 \, \left( \frac{ M_{\rm p} }{6M_{\oplus }} \right)^{2}\left( \frac{ R_{\rm core} }{1.6R_{\oplus }} \right)^{-4} \left( \frac{\zeta }{0.66} \right)\left(\frac{x_{\rm H_{2}O}}{10^{-2}}\right)^{-0.26} \left( \frac{\mu_{\rm atm}}{0.1\mu_{\rm H_{2}O}} \right)\left(\frac{P_{\rm BOA}}{\rm 10000~bar}\right)^{-0.26}. 
\label{eq:H2Osol}
\end{eqnarray}

In Equation (\ref{eq:H2sol}), the term $\phi_{\rm H_{2}}(0.47)^{P_{\rm BOA}[{\rm GPa}]}$ is close to 1 when the pressure is less than a few thousand bars. At high pressures, it exhibits temperature dependence and ranges from $\sim 1.5$ (2000 K) to $\sim 0.4$ (4000 K) at 30000 bar. However, we note that the range of $\phi_{\rm H_{2}}(0.47)^{P_{\rm BOA}[{\rm GPa}]}$ is relatively narrow and does not affect the result significantly.

If we ignore the He fraction in atmosphere, the conservation law of H can be expressed as follows:
%%%%%%%%%%%%%%%%%%%%
\begin{eqnarray}
N_{\rm H} %&=& 2 \{ ( 1 + r_{\rm H2} ) N_{\rm H2} + ( 1 + r_{\rm H2O} ) N_{\rm H2O} \} 
&=& \left\{ ( 1 + r_{\rm H_{2}} ) ( 1 - x_{\rm H_{2}O} ) +  ( 1 + r_{\rm H_{2}O} ) x_{\rm H_{2}O} \right\} N_{\rm atm,H}
\end{eqnarray}
%%%%%%%%%%%%%%%%%%%%
where $N_{\rm atm,H}$ is the total number of H in the atmosphere. As previously discussed in Section \ref{ss:model_dissolution}, the solubility of C-bearing species, except for the highly oxidized atmosphere case, is negligible. Ignoring the solubility of C-bearing species yields:
%%%%%%%%%%%%%%%%%%%%
\begin{eqnarray}
\left( \frac{\rm C}{\rm H} \right)_{\rm atm} &\sim & \frac{N_{\rm C}}{N_{\rm atm,H}} \\
&\sim & \left\{ ( 1 + r_{\rm H_{2}} ) ( 1 - x_{\rm H_{2}O} ) +  ( 1 + r_{\rm H_{2}O} ) x_{\rm H_{2}O} \right\} \left( \frac{\rm C}{\rm H} \right)_{\rm total}. 
\label{eq:CHrelation_basiceq}
\end{eqnarray}

When $x_{\rm H_{2}O} \ll 1$, this reduces to 
%%%%%%%%%%%%%%%%%%%%
\begin{eqnarray}
\left( \frac{\rm C}{\rm H} \right)_{\rm atm} &\sim & (1 + r_{\rm H_{2}}) \left( \frac{\rm C}{\rm H} \right)_{\rm total}.
\label{eq:CH_ratio_byH2}
\end{eqnarray}
%%%%%%%%%%%%%%%%%%%%

When combined with Equation (\ref{eq:H2sol}) and $\mu_{\rm m}\sim 0.06~\rm kg/mol$, it becomes apparent that the dissolution of \HH\ can increase $\left( \frac{\rm C}{\rm H} \right)_{\rm atm}$ approximately 1.6 to 8.0 times, depending on the surface conditions \citep[see also][]{2018ApJ...854...21C}.

On the other hand, if the second term is dominant,
%%%%%%%%%%%%%%%%%%%%
\begin{eqnarray}
\left( \frac{\rm C}{\rm H} \right)_{\rm atm} &\sim & r_{\rm H_{2}O} x_{\rm H_{2}O} \left( \frac{\rm C}{\rm H} \right)_{\rm total} \\
&\sim& 0.22 \, \left( \frac{ M_{\rm p} }{6M_{\oplus }} \right)^{2} \left( \frac{ R_{\rm core} }{1.6M_{\oplus }} \right)^{-4} \left( \frac{\zeta }{0.66} \right) \left( \frac{P_{\rm BOA}}{10000\,{\rm bar}} \right)^{-0.26} \left( \frac{\mu_{\rm atm}}{0.1\mu_{\rm H_{2}O}} \right) \left( \frac{x_{\rm H_{2}O}}{10^{-2}} \right)^{0.74} \left( \frac{\rm C}{\rm H} \right)_{\rm total}. 
\label{eq:CH_ratio_byH2O}
\end{eqnarray}
%%%%%%%%%%%%%%%%%%%%

$\left( \frac{\rm C}{\rm H} \right)_{\rm atm}$ increases as $x_{\rm H_{2}O}^{0.74}$ where the power-law index comes from the solubility law of \HHO. 

The condition when the second term becomes dominant can be estimated by taking the ratio between the two terms on the right-hand side of Equation (\ref{eq:CHrelation_basiceq}), approximating that $r_{\rm H_{2}} \gg 1$ and $r_{\rm H_{2}O} \gg 1$:
%%%%%%%%%%%%%%%%%%%%
\begin{eqnarray}
\frac{\rm 2nd\,term}{\rm 1st\,term} &=& \frac{( 1 + r_{\rm H_{2}O} ) x_{\rm H_{2}O}}{ ( 1 + r_{\rm H_{2}} ) ( 1 - x_{\rm H_{2}O} )} \\
&\sim & 16 \frac{1}{0.46^{P({\rm GPa})}\phi_{\rm H_{2}}}  \left(\frac{x_{\rm H_{2}O}}{1-x_{\rm H_{2}O}} \right) \left( \frac{P_{\rm BOA}}{10000\,{\rm bar}} \right)^{-0.26}\left( \frac{x_{\rm H_{2}O}}{10^{-2}} \right)^{-0.26} \\
&\sim & 0.16 \frac{1}{0.46^{P({\rm GPa})}\phi_{\rm H_{2}}} \left( \frac{P_{\rm BOA}}{10000\,{\rm bar}} \right)^{-0.26}\left( \frac{x_{\rm H_{2}O}}{10^{-2}} \right)^{0.74}.
\end{eqnarray}
%%%%%%%%%%%%%%%%%%%%

If we let $0.46^{P({\rm GPa})}\phi_{\rm H_{2}} \sim 1$, the second term becomes dominant when
%%%%%%%%%%%%%%%%%%%%
\begin{eqnarray}
\left( \frac{P_{\rm BOA}}{10000\,{\rm bar}} \right)^{-0.26}\left( \frac{x_{\rm H_{2}O}}{10^{-2}} \right)^{0.74}\gtrsim 60.  
\label{eq:relativeeffectH2OH2}
\end{eqnarray}
%%%%%%%%%%%%%%%%%%%%

Therefore, the C/H ratio can increase significantly, by more than several tens of times, in a low-pressure, oxidizing atmosphere case. This effect is illustrated by the red line in the upper panel of Figure \ref{fig:atmos_Oini_linear_surfacedepen}, which represents the 1000~bar surface pressure case, showing the highly enriched C/H ratio within the oxidized region that the Ferric-Ferrous governs.  

Equations (\ref{eq:H2sol}), (\ref{eq:CH_ratio_byH2}), and (\ref{eq:CH_ratio_byH2O}) suggest that the atmospheric C/O ratio would decrease as the atmosphere becomes more \HHO-rich, as the increase of the atmospheric C/H ratio is not as significant as that of \HHO. If \HH\ dissolution dominates the change in atmospheric C/H ratio and we assume the amounts of C-bearing species and He is negligible, the atmospheric C/O ratio can be approximated as follows, based on Equation (\ref{eq:CH_ratio_byH2}):

%%%%%%%%%%%%%%%%%%%%
\begin{eqnarray}
\begin{split}
\left( \frac{\rm C}{\rm O} \right)_{\rm atm} 
\sim \left\{ 1+4.6\phi_{\rm H_{2}}(0.47)^{P_{\rm BOA}[{\rm GPa}]}\left( \frac{\zeta }{0.66} \right)\left( \frac{ M_{\rm p} }{6M_{\oplus }} \right)^{2}\left( \frac{ R_{\rm core} }{1.6R_{\oplus }} \right)^{-4}\left( \frac{\mu_{\rm atm}}{0.1 \mu_{\rm m}} \right) \right\} \left( \frac{\rm O}{\rm H} \right)_{\rm atm}^{-1} \left( \frac{\rm C}{\rm H} \right)_{\rm total}\\
\sim \left\{ 200+920\phi_{\rm H_{2}}(0.47)^{P_{\rm BOA}[{\rm GPa}]}\left( \frac{\zeta }{0.66} \right)\left( \frac{ M_{\rm p} }{6M_{\oplus }} \right)^{2}\left( \frac{ R_{\rm core} }{1.6R_{\oplus }} \right)^{-4} \left( \frac{\mu_{\rm atm}}{0.1 \mu_{\rm m}} \right) \right\} \left( \frac{x_{\rm H_{2}O}}{10^{-2}} \right)^{-1} \left( \frac{\rm O}{\rm H} \right)_{\rm accreted} \left( \frac{\rm C}{\rm O} \right)_{\rm accreted}\\
\sim \left\{ 0.12+0.53\phi_{\rm H_{2}}(0.47)^{P_{\rm BOA}[{\rm GPa}]}\left( \frac{\zeta }{0.66} \right)\left( \frac{ M_{\rm p} }{6M_{\oplus }} \right)^{2}\left( \frac{ R_{\rm core} }{1.6R_{\oplus }} \right)^{-4} \left( \frac{\mu_{\rm atm}}{0.1 \mu_{\rm m}} \right) \right\} \left( \frac{x_{\rm H_{2}O}}{10^{-2}} \right)^{-1} \left( \frac{\rm C}{\rm O} \right)_{\rm accreted}
\label{eq:CO_ratio_byH2}
\end{split}
\end{eqnarray}
%%%%%%%%%%%%%%%%%%%%
where $\left( \frac{\rm O}{\rm H} \right)_{\rm accreted}$ is the O/H ratio of accreted volatiles, which has a value of $5.8\times10^{-4}$ in this study. It is evident that the C/O ratio follows a simple inverse proportional relation with $x_{\rm H_{2}O}$. If $x_{\rm H_{2}O}$ exceeds a few percent, the C/O ratio diminishes to 10\% of that of the assumed composition of accreted gas.

On the other hand, if the \HHO\ dissolution governs the change in atmospheric C/H, the atmospheric C/O follows the relation as follows:
%%%%%%%%%%%%%%%%%%%%
\begin{eqnarray}
\begin{split}
\left( \frac{\rm C}{\rm O} \right)_{\rm atm} 
\sim 0.22\left( \frac{\zeta }{0.66} \right)\left( \frac{ M_{\rm p} }{6M_{\oplus }} \right)^{2}\left( \frac{ R_{\rm core} }{1.6R_{\oplus }} \right)^{-4}\left( \frac{P_{\rm BOA}}{10000\,{\rm bar}} \right)^{-0.26}\left( \frac{\mu_{\rm atm}}{0.1\mu_{\rm H_{2}O}} \right)\left( \frac{x_{\rm H_{2}O}}{10^{-2}} \right)^{0.74}\left( \frac{\rm O}{\rm H} \right)_{\rm atm}^{-1} \left( \frac{\rm C}{\rm H} \right)_{\rm total}\\
\sim 44\left( \frac{\zeta }{0.66} \right)\left( \frac{ M_{\rm p} }{6M_{\oplus }} \right)^{2}\left( \frac{ R_{\rm core} }{1.6R_{\oplus }} \right)^{-4}\left( \frac{P_{\rm BOA}}{10000\,{\rm bar}} \right)^{-0.26}\left( \frac{\mu_{\rm atm}}{0.1\mu_{\rm H_{2}O}} \right)\left( \frac{x_{\rm H_{2}O}}{10^{-2}} \right)^{-0.26}\left( \frac{\rm O}{\rm H} \right)_{\rm accreted} \left( \frac{\rm C}{\rm O} \right)_{\rm accreted}\\
\sim 0.026\left( \frac{\zeta }{0.66} \right)\left( \frac{ M_{\rm p} }{6M_{\oplus }} \right)^{2}\left( \frac{ R_{\rm core} }{1.6R_{\oplus }} \right)^{-4}\left( \frac{P_{\rm BOA}}{10000\,{\rm bar}} \right)^{-0.26}\left( \frac{\mu_{\rm atm}}{0.1\mu_{\rm H_{2}O}} \right)\left( \frac{x_{\rm H_{2}O}}{10^{-2}} \right)^{-0.26}\left( \frac{\rm C}{\rm O} \right)_{\rm accreted}.
\label{eq:CO_ratio_byH2O}
\end{split}
\end{eqnarray}
%%%%%%%%%%%%%%%%%%%%

Equation (\ref{eq:CO_ratio_byH2O}) demonstrates that when the dissolution of \HHO\ plays a dominant role in C/H enrichment, the relationship between C/O depletion and $x_{\rm H_{2}O}$ is milder ($\propto x_{\rm H_{2}O}^{-1} \rightarrow x_{\rm H_{2}O}^{-0.26}$). 

%%%%%%%%%%%%%%%%%%%%
\begin{figure*}
\centering
\includegraphics[width=18cm]{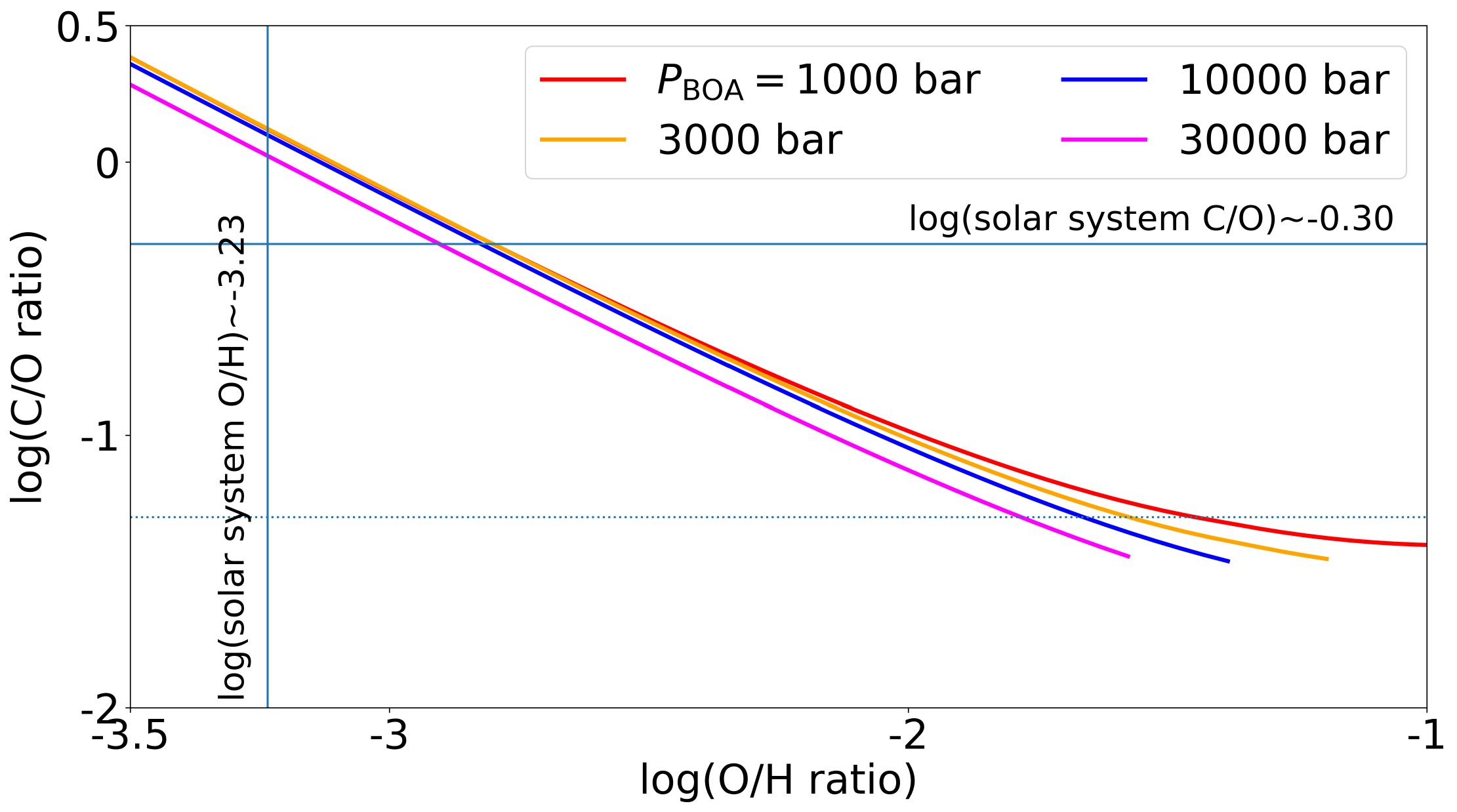} 
\caption{The relationship between atmospheric O/H and C/O ratios and their dependence on surface pressure. Different colors represent different surface pressures: 1000, 3000, 10000, and 30000 bar. The blue vertical and horizontal lines denote the solar system values of O/H and C/O. A dotted horizontal line marks 10\% of the solar C/O.}
\label{fig:OHCOrelation}
\end{figure*} 
%%%%%%%%%%%%%%%%%%%%

Figure \ref{fig:OHCOrelation} shows the O/H--C/O relationship and its dependence on surface pressure. As discussed earlier using the two C/O equations, the decrease in C/O begins with a monotonic trend, becomes gentler as $x_{\rm H_{2}O}$ increases, and finally reaches $\sim 10\%$ of the nebula gas value under an oxidized atmosphere with $x_{\rm H_{2}O} \ge$ a few percentages. We note that the C/O ratio decreases with increasing pressure due to the lower $r_{\rm H2}$ and $r_{\rm H2O}$, as Equation (\ref{eq:H2sol}) and Equation (\ref{eq:H2Osol}) show.

\subsection{Dependence on Fe/Si and size of magma}
\label{ss:atmospheric composition_magmaparamdepen}

In this section, we discuss two additional magma properties that affect the atmospheric composition beyond the magma redox state (iron speciation) already covered in Section \ref{ss:atmospheric composition}: the magma composition (Fe/Si ratio) and the reactive amount of magma (reactive mass fraction to the total magma mass, $\frac{\zeta}{0.66}$). Our model assumes fiducial parameters of a Fe/Si ratio of 0.17 (MORB-like) and $\frac{\zeta}{0.66}$ of 1. The ranges considered in this section vary from $10^{-1}$ to $10^{0.5}$ times the MORB value for Fe/Si and from $10^{-2}$ to 1 for $\frac{\zeta}{0.66}$. We assume that the $T_{\rm BOA}$ is 3000 K and the $P_{\rm BOA}$ is 10000 bar.

\begin{figure*}
\centering
\includegraphics[width=18cm]{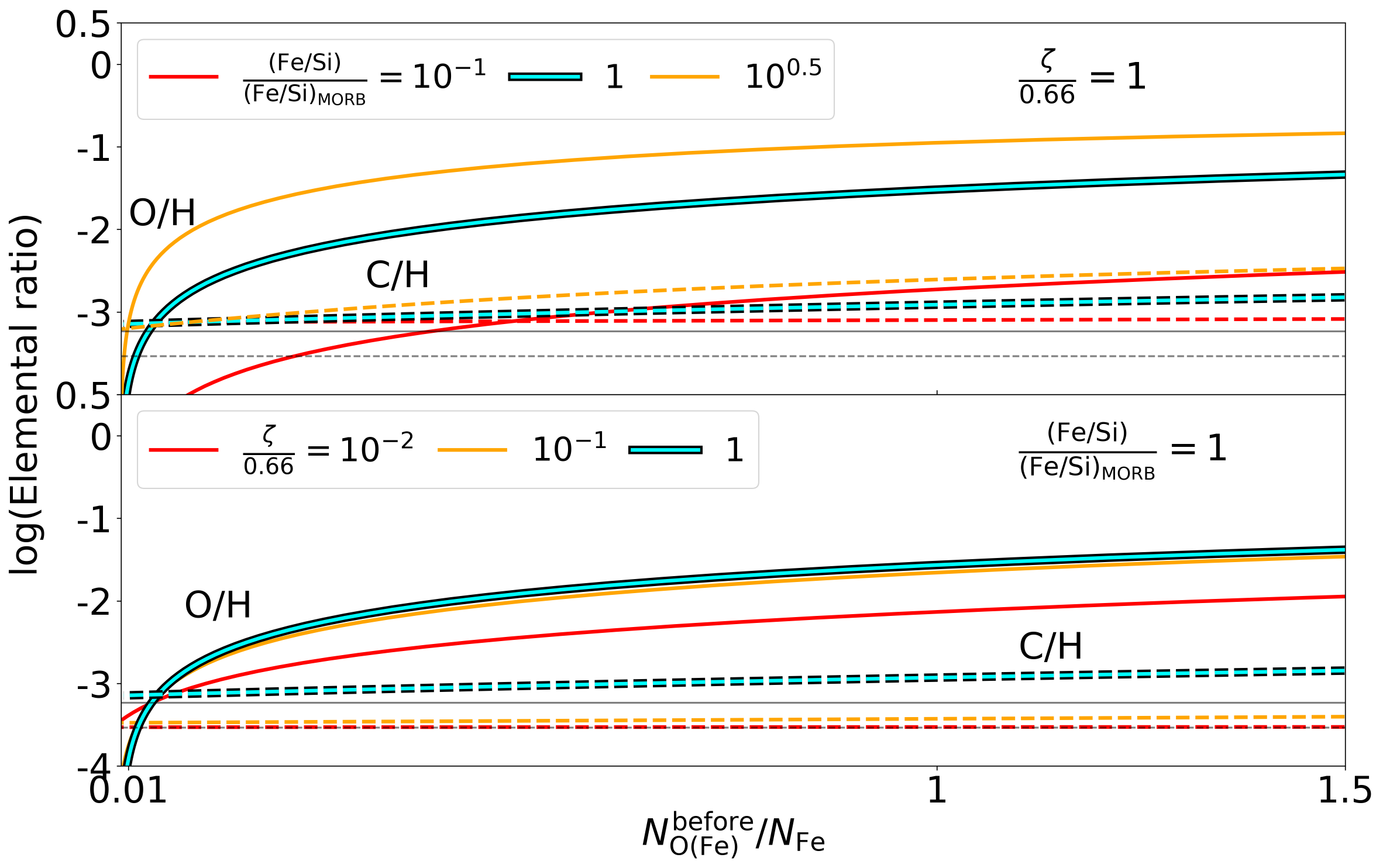}
\caption{The relation between the atmospheric H-C-O ratio and $N_{\rm O(Fe)}^{\rm before}$/$N_{\rm Fe}$, with its dependency on two magma properties, Fe/Si ratio and size of reactive magma ($\frac{\zeta}{0.66}$). (Upper) We show three Fe/Si cases: $10^{-1}$, $1$, and $10^{0.5}$ times the MORB value (0.17). The reactive magma fraction is fixed at 1. (Lower) Three reactive magma fraction cases are shown: $10^{-2}$, $10^{-1}$, $1$. The Fe/Si is fixed at the MORB value. The surface temperature and pressure are fixed at 3000 K and 10000 bar, respectively.}
\label{fig:atmos_Oini_linear_magmadepen}
\end{figure*}

Figure \ref{fig:atmos_Oini_linear_magmadepen} illustrates the relationship between the atmospheric O/H and C/H ratios and the magma iron speciation (redox state), focusing on the dependence on magma properties. The upper panel shows that iron-poor magma corresponds to a smaller O/H ratio for the same iron speciation, which is expected due to the limited amount of oxidized Fe species. This result is also consistent with the atmospheric composition trends with respect to different FeO content \cite[e.g.,][]{2020ApJ...891..111K}. 

%%%%%%%%%%%%%%%%%%%%
%%%%%%%%%%%%%%%%%%%%

The lower panel demonstrates that a smaller effective magma fraction results in a lower O/H ratio when $\frac{\zeta}{0.66} <0.1$. This can be understood based on the discussion in Section \ref{ss:atmospheric composition}. As long as we can ignore the second term of Equation (\ref{eq:O_conservation}), the $\zeta $-dependence of O/H is minor. When $\zeta $ becomes smaller, however, the second term can no longer be ignored, and Equation (\ref{eq:conservation_O_modified}) becomes

\begin{equation}
\frac{N_{\rm O(Fe)}^{\rm before}}{N_{\rm Fe}} f_{\rm Fe} = a_1 x_{\rm H_{2}O} + \frac{N_{\rm atm}}{N_{\rm m}} x_{\rm H_{2}O} + a_2  x_{\rm H_{2}O}^{0.74}.
\end{equation}

Now, for a given $N_{\rm O(Fe)}^{\rm before}$/$N_{\rm Fe}$, the smaller $N_{\rm m}$ is compensated by a lower $x_{\rm H2O}$. 

Note that the atmospheric O/H-C/O relation remains consistent across different Fe/Si cases. However, a smaller reactive magma amount causes the same O/H ratio to correspond to a lower C/O ratio, as can also be inferred from Equation (\ref{eq:CO_ratio_byH2}) and Equation (\ref{eq:CO_ratio_byH2O}).

\subsection{Dependence on atmospheric parameters}
\label{ss:atmospheric composition_planetaryparamdepen}

In our model, the two key factors that control the atmospheric composition in contact with magma ocean, $P_{\rm BOA}$ and $T_{\rm BOA}$, are related to the observable parameters $R_{\rm p}$, $T_\mathrm{eq}$, and $M_{\rm p}$ through the assumption on $P_{\rm RCB}$ and Earth-like rocky core. In this section, we show the atmospheric composition as a function of these parameters. The fiducial parameters are $M_{\rm p}=6M_{\oplus }$, $T_{\rm eq}=750$~K, and $P_{\rm RCB}=10$~bar. Instead of $R_{\rm p}$, we use the atmospheric thickness, $\Delta R = R_{\rm p} - R_{\rm core}$, to represent the planetary radius. The fiducial value for $\Delta R$ is $0.4R_{\rm \oplus}$. Note that under the assumption of an Earth-like rocky core, the core radius is found to be $R_{\rm core}=1.6R_{\rm \oplus}$, and thus $\Delta R=0.4R_{\oplus }$ corresponds to $R_{\rm p}=2.0R_{\oplus }$. We change one of these parameters at a time to clarify the dependence of each parameter.
The ranges of parameters considered are 0.1 to 0.8 $R_\oplus$ $(\rm \Delta R)$, 550 to 1150 $\rm K$ $(T_{\rm eq})$, 0.1 to 100 $\rm bar$ $(P_{\rm RCB})$, and 2 to 8 $M_{\rm \oplus}$ $(M_{\rm p})$.

\begin{figure*}
\centering
\includegraphics[width=18cm]{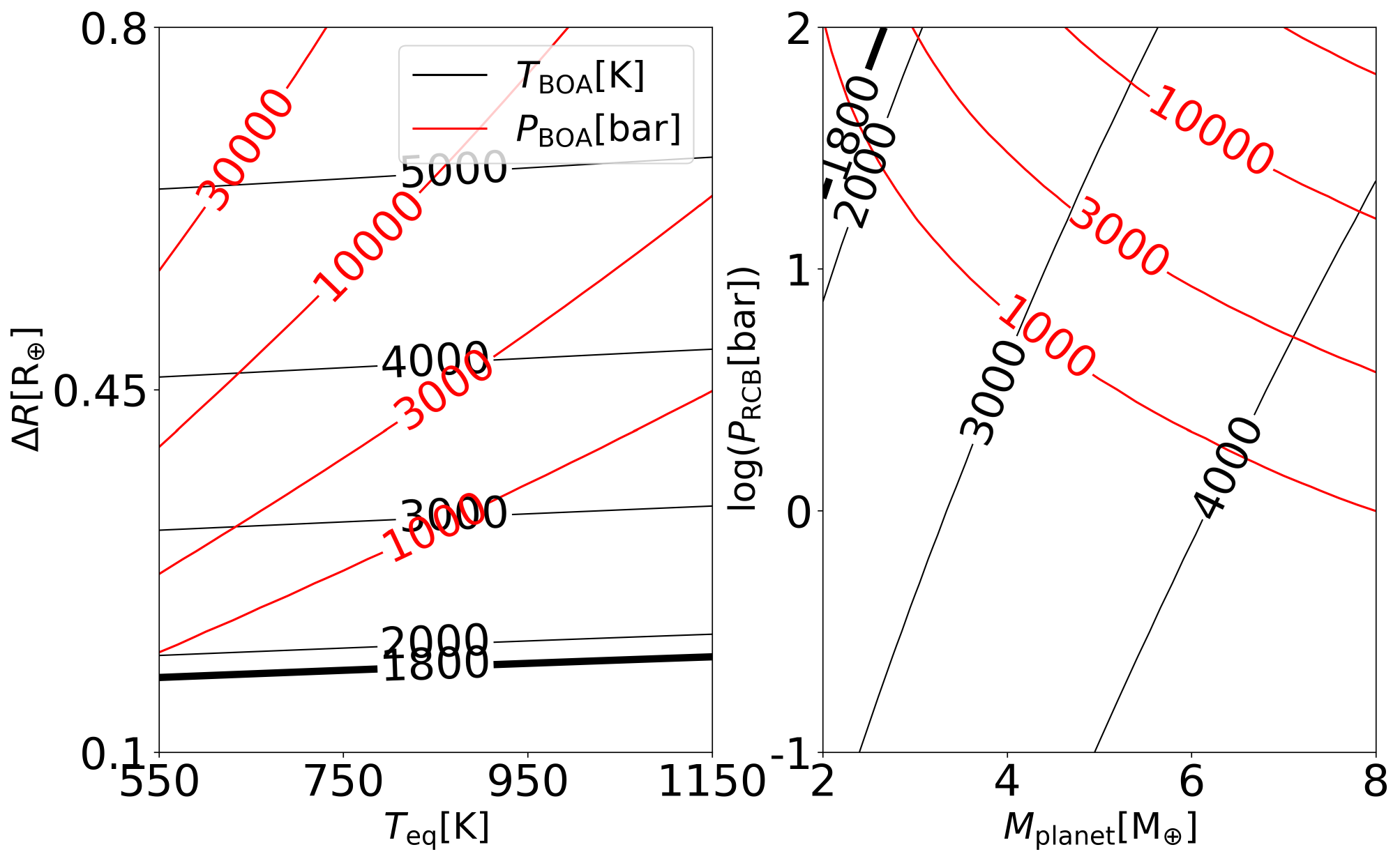} 
\caption{The relationship among $P_{\rm BOA}$ (red contour), $T_{\rm BOA}$ (black contour), $\Delta R$, $T_{\rm eq}$, $P_{\rm RCB}$, and $M_{\rm planet}$ in our model. The atmosphere is composed of \HH\ (82\%), He (17\%), and \HHO\ (1\%). In the left panel, $P_{\rm RCB}$ and $M_{\rm planet}$ are fixed at $10$~bar and $6M_{\oplus}$, respectively. In the right panel, $\Delta R$ and $T_{\rm eq}$ are fixed at $0.4R_{\rm \oplus}$ and $750$~K, respectively.}
\label{fig:surfaceconditions}
\end{figure*}

Figure \ref{fig:surfaceconditions} illustrates the ranges of $P_{\rm BOA}$ and $T_{\rm BOA}$ under an oxidized atmosphere and their dependencies on the planetary parameters. A thicker atmosphere (larger $\Delta R$), a colder upper atmosphere (smaller $T_\mathrm{eq}$), and a heavier planetary mass (larger $M_{\rm p}$) correspond to larger $P_{\rm BOA}$ and $T_{\rm BOA}$. On the other hand, the effects of $P_\mathrm{RCB}$ are different for pressure and temperature: the larger $P_{\rm RCB}$ corresponds to higher $P_{\rm BOA}$ and lower $T_{\rm BOA}$.

%%%%%%%%%%%%%%%%%%%%
%%%%%%%%%%%%%%%%%%%%

\begin{figure*}
\centering
\includegraphics[width=18cm]{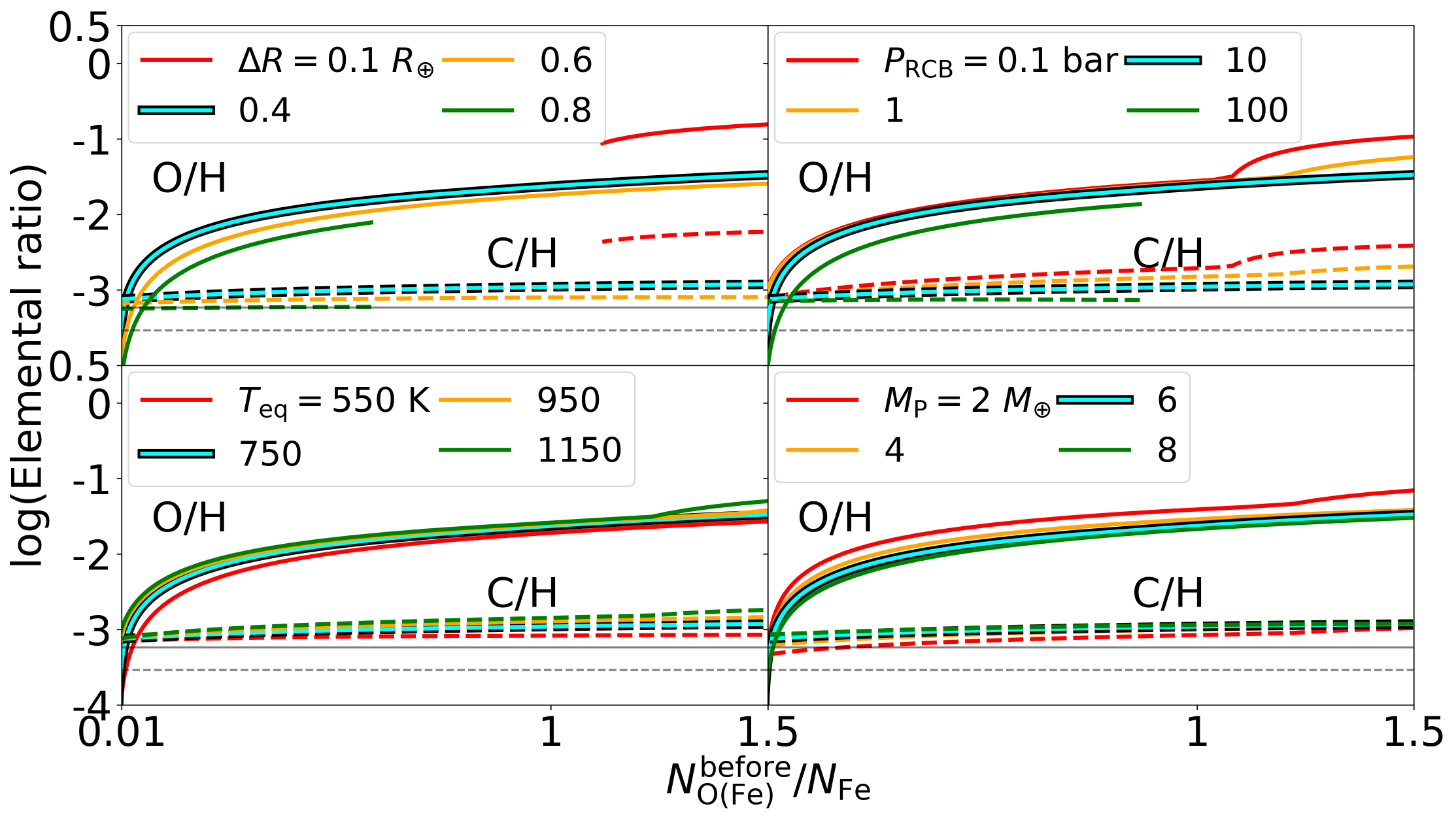}
\caption{The relation between the atmospheric H-C-O ratio and $N_{\rm O(Fe)}^{\rm before}$/$N_{\rm Fe}$, with respect to four planetary parameters: ($\Delta R$, $T_{\rm eq}$, $P_{\rm RCB}$, and $M_{\rm planet}$). The x-axis denotes $N_{\rm O(Fe)}^{\rm before}$/$N_{\rm Fe}$, while the y-axis represents atmospheric composition. Solid lines represent O/H ratio, while dashed lines represent C/H ratio. Horizontal solid and dashed lines indicate $\rm \log(O/H)$ and $\rm \log(C/H)$ of accreted volatiles, which are approximately $-3.23$ and $-3.53$, respectively. (Upper left) Four different $\Delta R $ cases: 0.1, 0.4, 0.6, and 0.8 $(R_{\oplus})$. (Upper right) Four $P_{\rm RCB}$ cases: 0.1, 1, 10, and 100 bar. (Lower left) Four $T_{\rm eq}$ results: 550, 750, 950, and 1150~K. (Lower right) Four $M_{\rm planet}$ lines: 2, 4, 6, and 8 $M_{\oplus}$. Cyan lines serve as the fiducial values: $\Delta R = 0.4 R_{\oplus}$, $P_{\rm RCB} = 10$~bar, $T_{\rm eq} = 750$~K, and $M_{\rm planet} = 6M_{\oplus}$. Data for cases where $T_{\rm BOA}$ and $P_{\rm BOA}$ are out of the parameter range of this study are not shown.}
\label{fig:atmos_Oini_linear}
\end{figure*}

We show the relation between the atmospheric composition (O/H and C/H) and the magma redox state and its dependence on the planetary parameters in Figure \ref{fig:atmos_Oini_linear}. 
It demonstrates that atmospheres with smaller $\Delta R$, $M_{\rm p}$, and $P_\mathrm{RCB}$ and larger $T_\mathrm{eq}$ result in larger O/H and correspondingly elevated C/H ratios, consistent with the expectations from the $P_{\rm BOA}$ result shown in Figure \ref{fig:surfaceconditions}. From these investigations, we find that the effect of the planetary parameters, except for $\Delta R$ and $P_\mathrm{RCB}$, is almost negligible. However, O/H is sensitive to small $\Delta R$ and small $P_\mathrm{RCB}$ since both can make a small atmospheric pressure.

%%%%%%%%%%%%%%%%%%%%
%%%%%%%%%%%%%%%%%%%%

\subsection{Effect of the magma solidification}
\label{ss:effectofmoltenamount}

In this section, we demonstrate the effect of magma solidification on the atmospheric composition following the methodology described in Section {\ref{ss:magma_solidification}}. We assume that the $P_{\rm BOA}$ before the solidification is 5000 bar and the $T_{\rm BOA}$ remains constant at 3000 K. Note that fixing the BOA temperature does not cause significant error in estimating the atmospheric composition (see Section \ref{ss:atmospheric composition}).

\begin{figure}
\centering
\includegraphics[width=18cm]{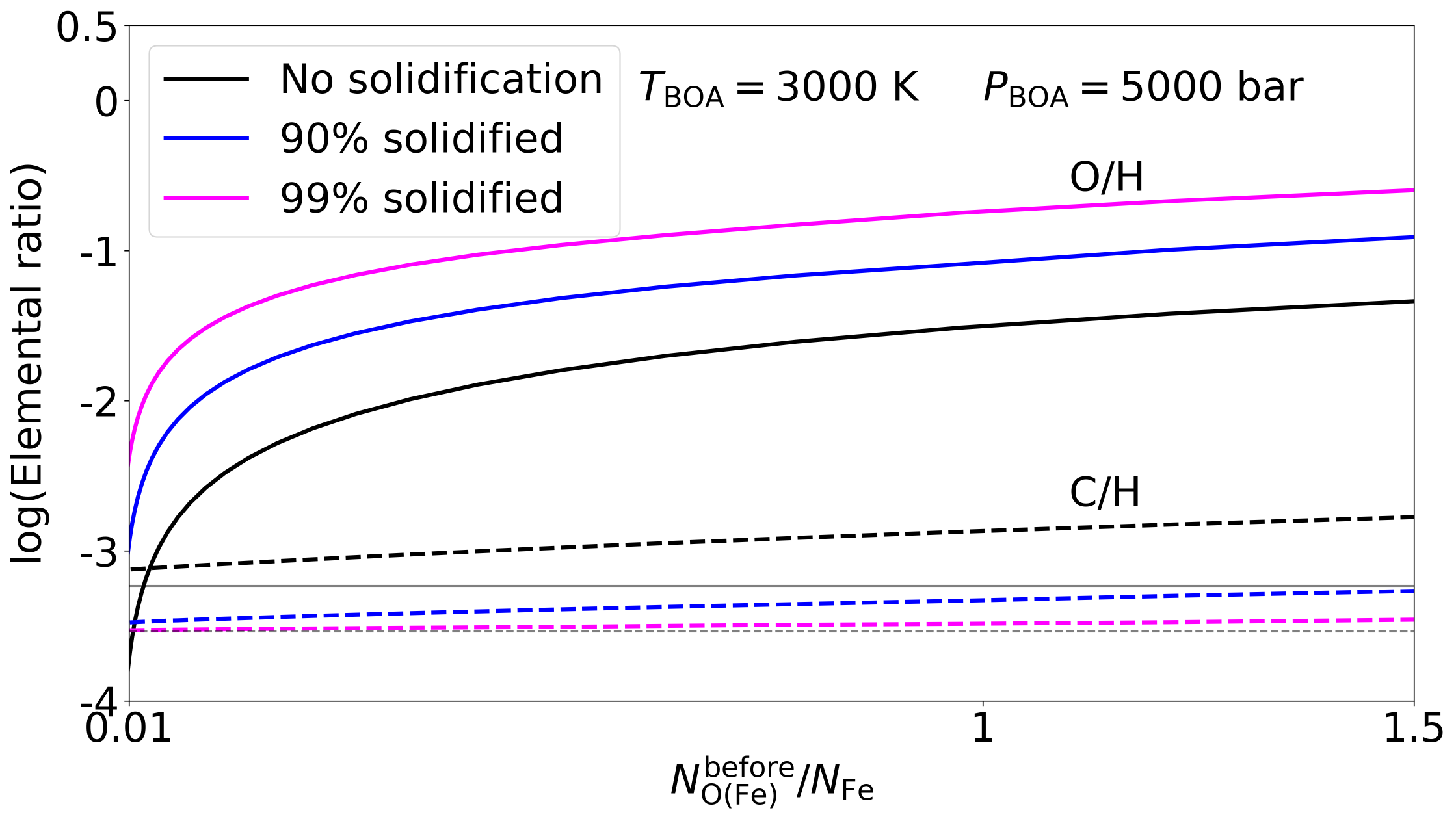} 
\caption{The relation between the atmospheric H-C-O ratio and silicate redox state (before the reaction), with its dependency on the molten fraction of silicate core. The variable for the x-axis is $N_{\rm O(Fe)}^{\rm before}$/$N_{\rm Fe}$, which denotes the redox state of the silicate before the reaction. The y-axis corresponds to the elemental ratio on a logarithmic scale. The solid line corresponds to the O/H ratio, while the dashed line represents the C/H ratio. $\rm log(O/H)$ and $\rm log(C/H)$ of accreted volatiles are approximately $-3.23$ and $-3.53$, respectively. The lines of different colors represent various mass fractions of magma ($\zeta$), with $\zeta = 6.6 \times 10^{-1}$ depicted in black, $\zeta = 6.6 \times 10^{-2}$ in blue, and $\zeta = 6.6\times 10^{-3}$ in magenta.}
\label{fig:solidification}
\end{figure}

Figure \ref{fig:solidification} shows the case with $\frac{\zeta}{0.66} = 10^{-1}$, $10^{-2}$, and $10^{-3}$ scenarios. It is clear that the $x_{\rm H2O}$ increases under the same $N_{\rm O(Fe)}^{\rm before}$/$N_{\rm Fe}$ due to the extraction of dissolved \HHO\ from the solidified rock layer to the upper molten magma and atmosphere. At the same time, the C/H ratio decreases due to the decrease of $\zeta$. These two effects imply that the C/O ratio can decrease further with the magma solidification effect.

%%%%%%%%%%%%%%%%%%%%
%%%%%%%%%%%%%%%%%%%%

\section{Discussion}

\subsection{Implications for the internal structure and formation scenarios}

We have examined the range of atmospheric composition expected for the scenario where the rocky planetary core accreted the nebula gas. Can such atmospheric composition be a useful clue to the formation pathways of sub-Neptunes? To address this question, we investigate the expected volatile composition of sub-Neptunes under an alternative formation scenario: the assembly of icy planetesimals outside of the snow line (referred to as ice-rich sub-Neptunes hereafter) and compare it to the case of magma-bearing sub-Neptunes. 

The composition of ice-rich sub-Neptunes would be determined from the elemental composition of the solid components in the protoplanetary disk (e.g., \citealt{2011ApJ...743L..16O}). The elemental abundance ratio of gaseous and solid components within the protoplanetary disk is expected to vary as a function of the orbital radius relative to the sublimation lines of volatile species, although there are ongoing debates on the underlying mechanisms and chemistry of the protoplanetary disk (e.g., \citealt{2023ARA&A..61..287O}). 
For example, the C/O ratio of solids (gas) beyond the \HHO\ snow line would be smaller (larger) than that interior of the \HHO\ snow line. The exact C/O value in solids at a given location would depend on the ratio of carbon in volatile form to refractory form. Under the typical scenario where about 75\% of the carbon exists in volatile form and a rock-to-water ratio of approximately 1, the C/O in solids beyond the \HHO\ snow line would be a few tens of percent of the nebula value. If the planetesimals beyond the \COO\ and \CO\ snow lines also accrete, the planetary C/O would further increase. In these cases, we anticipate that the oxidized atmosphere in contact with magma ocean, which exhibits a depleted C/O ratio well below 10\% of nebula value, would be distinct from the expectation for the ice-rich sub-Neptune scenario. 

However, uncertainties exist in the previously described simplistic scenario. In particular, the fraction of refractory carbon could be smaller due to the irreversible destruction processes (e.g., \citealt{2021SciA....7.3632L}), which would reduce C/O in ice-rich sub-Neptune scenarios to an unknown extent. This uncertainty might overlap with the C/O expected in the magma scenario considered in this paper. 
Further studies on the plausible range of the volatile compositions of planetesimals beyond the \HHO\ snow line and the formation pathways of small planets (e.g., \citealt{2022NatAs...6.1296K}) are needed. These studies will help to clarify further the atmospheric compositions that can distinguish the formation pathways of sub-Neptunes, and this is left for future work. 

\subsection{Implications for observation}
\label{ss:implicationobs}

In this section, we discuss the spectral features of the atmospheric scenarios we find. Using the open-source code TauREx3 (\citealt{2021ApJ...917...37A}), we simulate the transmission spectra for two representative atmospheric scenarios from our main results (see Figure \ref{fig:OHCOrelation}): (1) the reduced case with $\log _{10}{\rm O/H}=-3.5$ and $\log _{10}{\rm C/O}=0.3$, and (2) the oxidized case with $\log _{10}{\rm O/H}=-1.5$ and $\log _{10}{\rm C/O}=-1.3$. 

The atmospheric temperature-pressure structure is calculated with the nominal assumption: $\Delta R=0.4~\rm R_\oplus$, $T_{\rm eq}=750~\rm K$, $P_{\rm RCB}=10~\rm bar$, and $M_{\rm p} = 6~\rm M_{\oplus}$. To compute the molecular composition of the atmosphere, we assume that molecular composition is constant above the quenching point, where the vertical mixing timescale overwhelms the chemical timescale (e.g., \citealt{2014ApJ...784...63H}). We approximately represent this quench point by a single pressure. Noting that, in reality, the quenching point can depend on the molecules and the background atmospheric composition, we explore the dependence on the quenching pressure by varying it between 100 and 10~bar. We also tested the case where the quenching pressure is located at 1 bar, but it does not differ significantly from the 10 bar case. In all of these cases, we find that O and C in the atmosphere exist predominantly as \HHO\ and \CHHHH\, respectively \citep[see Figure 11 of][]{2014ApJ...784...63H}. Therefore, the mixing ratio of \HHO\ and \CHHHH\ is simply $\sim 2\log ({\rm O}/{\rm H})$ and $\sim \log ({\rm C}/{\rm H})$, respectively.

\begin{figure}
\includegraphics[width=18cm]{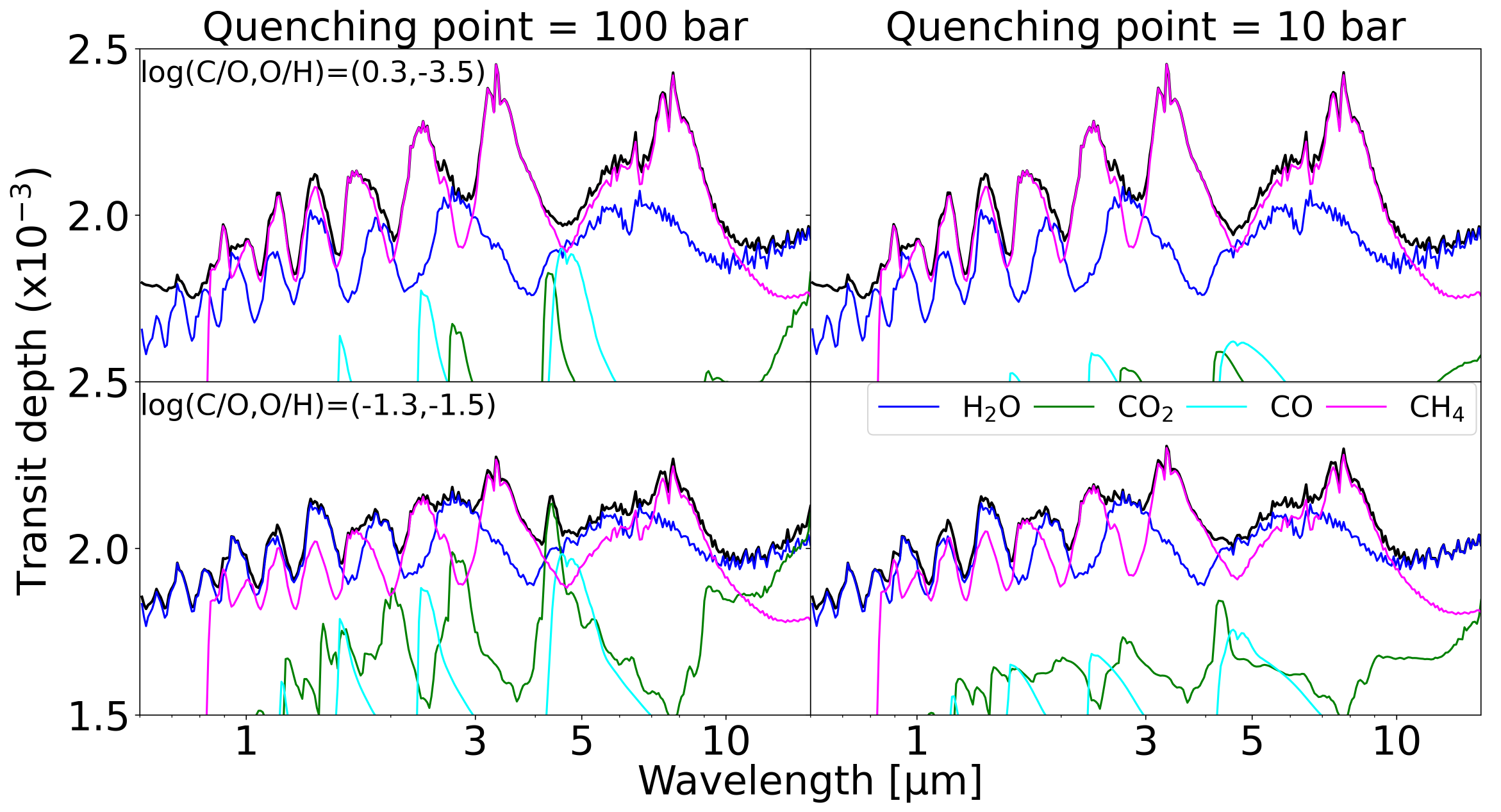} 
\caption{Simulated transmission spectra of two representative atmospheric cases from our results (reduced, C-enriched one and oxidized, C-poor one) with varying quenching pressures. The top row corresponds to the reduced atmosphere (atmospheric O/H ratio of $10^{-3.5}$ and C/O ratio of $10^{0.3}$), while the bottom row illustrates the oxidized atmosphere (atmospheric O/H ratio of $10^{-1.5}$ and C/O ratio of $10^{-1.3}$). Different colored lines represent the contributions of different species: \HHO\ (blue), \CO\ (cyan), \COO\ (green), and \CHHHH\ (magenta). The left column shows the case with a quenching point at 100 bar, while the right column denotes the shallower quenching point at 10 bar.}
\label{fig:spectrum}
\end{figure}

Figure \ref{fig:spectrum} illustrates the transmission spectra with a spectral resolution of 100 for the two atmospheric scenarios we mentioned above. The left and right columns correspond to two different quenching points. We assumed that the parent star is an M-type star with a temperature of 3600~K and a stellar radius of $R_{\star} =  0.42R_{\odot}$. If one wants to consider the transit spectra for the planets accompanied by other stars, the transit depth can be scaled by $R_{\star}^{-2}$. All spectra are primarily shaped by the absorption bands of \HHO\ and \CHHHH. The dominant features for the reduced scenario come from \CHHHH\ (1.65 $\mu $m, 2.4 $\mu $m, 3.3 $\mu $m, and 6.15 $\mu $m), while the \HHO\ features are also prominent (1.9 $\mu $m and 2.87 $\mu $m) in the oxidized case. Although the absorption bands of \HHO\ and \CHHHH\ tend to overlap at shorter wavelengths, they diverge at longer wavelengths, in particular beyond 2$\mu $m, enabling the detection of both species with high S/N observations. Notably, the oxidized atmosphere exhibits a weak signal of \COO\ at 4.3 $\mu $m in cases where the quenching point is as deep as 100 bar, providing additional constraints on O/H. 
Through the strong spectral features of \CHHHH\ and \HHO, we anticipate that finding the atmospheric H-C-O ratio from the spectral observation would be straightforward. Additionally, the mean molecular weight, another observable parameter \citep[e.g.,][]{2023ApJ...951...96G, 2024arXiv240303325B}, increases from 2.4 g/mol to 3.2 g/mol as the O/H ratio shifts from reduced to oxidized states.

%%%%%%%%%%%%%%%%%%%%
%%%%%%%%%%%%%%%%%%%%

Note that the signal strengths of the molecule spectral peaks in our demonstration (Figure \ref{fig:spectrum}) could be overestimated compared to transmission spectra shown in previous studies \citep[e.g.,][]{2019ApJ...877..109K}. 
This discrepancy arises because we did not consider photolysis caused by stellar UV irradiation, which can significantly reduce the abundance of molecules in low-pressure regions. Such inhomogeneity is prominent at altitudes where the timescale of photochemistry is shorter than the mixing timescale; for instance, \citet{2019ApJ...877..109K} shows that \HHO\ and \CHHHH\ can be fully photo-dissociated at pressures below $10^{-4.5}$ bar under intense UV irradiation ($10^5$ times that of GJ 1214). In our simulations, for the reduced and carbon-enriched case (upper left panel of Figure \ref{fig:spectrum}), we confirmed that complete photolysis under $10^{-4.5}$ bar would set the maximum transit depth to be $\sim 2.2 \times 10^{-3}$, reducing the spectral signal of \CHHHH\ and \HHO. Haze, which could photochemically form from \CHHHH, and refractory clouds that have not been considered here could further weaken the signals \citep[e.g.,][]{Ma+2023}. A detailed investigation of the impact of photochemistry and photochemical haze/clouds on transmission spectra will be our future study. 

\subsection{Comparison to the C-enrichment by atmospheric escape}
 
While we focused on the effect of a magma ocean on the atmospheric composition, another well-known factor that alters (or enriches) the atmospheric composition is atmospheric loss \cite[e.g.,][]{1981Icar...48..150W}. When the planetary upper atmosphere receives the intense XUV, \HH-rich atmospheres of sub-Neptunes undergo hydrodynamic escape, where individual atmospheric species escape to space at a rate that depends on the mass of the species. In this section, we estimate the change in atmospheric composition solely through such an atmospheric escape process and compare its effect against the effect of interaction with magma. Through this comparison, we aim to find the parameter space where the effect of magma can be identified.

We mainly follow the methodology applied in \citet{2016ApJ...829...63S}. The lightest atom, hydrogen, escapes from the atmosphere and drags heavier atoms (such as C and O) away. 
The escaping rate of heavier element ($\phi_{\rm 2}$) and the escaping rate of lighter element ($\phi_{\rm 1}$) satisfy the following equation (e.g., \citealt{1987Icar...69..532H}): 
\begin{equation}
  \phi_{2}=\begin{cases}
    \phi_{1}\frac{X_{\rm 2}}{X_{\rm 1}}\frac{\mu_{\rm c}-\mu_{\rm 2}}{\mu_{\rm c}-\mu_{\rm 1}}, & \text{if $\mu_{\rm 2}<\mu_{\rm c}$}\\
    0 & \text{if $\mu_{\rm 2}>\mu_{\rm c}$},
  \end{cases} 
\end{equation}
 where the ($X_{\rm 2}$ and $X_{\rm 1}$) are the molar concentration, and ($\mu_{\rm 2}$ and $\mu_{\rm 1}$) are the atmic mass of each species. The $\mu_{\rm c}$ is the crossover mass, determining the mass criterion for drag to occur. The crossover mass follows the equation: \begin{equation}
\mu_{\rm c}=\mu_{\rm 1}+\frac{k_{\rm B}T\phi_{\rm 1}}{b_{\rm 12}gX_{\rm 1}m_{\rm p}},
\end{equation} where $k_{\rm B}$ is the Boltzmann constant, $m_{\rm p}$ is the proton mass, and $b_{\rm 12}$ is the binary diffusion coefficient between two species. We refer to the neutral-neutral atom pair binary diffusion coefficient from \citet{1990Icar...84..502Z} and \citet{2022ApJ...934..137Y}, given by:
\begin{equation}
b_{ij}=1.96 \times 10^{6}\frac{T^{0.5}}{\mu_{ij}^{0.5}},
\end{equation} where $\mu_{ij}$ represents the reduced mass between species i and j. The escape flux without the heavier element (reference flux) is assumed to follow an energy-limited escape, satisfying the following equation (e.g., \citealt{1996JGR...10126039C, 2016ApJ...829...63S}): 
\begin{equation}
\phi_{\rm ref}=\frac{\epsilon F_{\rm XUV}r_{\rm p}}{4G\mu_{\rm 1}m_{\rm p}M_{\rm p}},
\end{equation} 
where $F_{\rm XUV}$ is the stellar XUV flux, $r_{\rm p}$ is the planetary radius, $M_{\rm p}$ is the planetary mass, and $\epsilon$ is the heating efficiency. The three escape rates ($\phi_{\rm ref}$, $\phi_{\rm 1}$, $\phi_{\rm 2}$) are related to each other: $\phi_{\rm ref}=\mu_{\rm 1}\phi_{\rm 1}+\mu_{\rm 2}\phi_{\rm 2}$.

The escape rate and crossover mass equations suggest that the escape rate of heavier elements is smaller than that of lighter elements, and the difference in escape rates increases as the escape rate of the lighter element decreases. Therefore, we anticipate that atmospheric escape will lead to an increase in the atmospheric O/H and C/H ratios and a slight decrease in the atmospheric C/O ratio. The selective escape effect is expected to be more pronounced under weaker XUV radiation and smaller atmospheric amounts, provided that the XUV radiation remains sufficiently strong to cause atmospheric escape.

Using the above equations, we calculate the enrichment of O/H and C/O under the following assumptions. First, we assume that the planet has evolved for 5 Gyr and finally has an atmosphere of the corresponding thickness and equilibrium temperature. Second, we consider an early M-type host star (M1-M2 type) with its XUV flux following the early M-type stellar evolution model proposed by \citet{2020ApJ...895....5P}. Third, we neglect the evolution of the planetary orbital radius and planetary radius. The orbital radius is translated to the planetary equilibrium temperature so as to match the parameterization of our magma-atmosphere model, with the assumption that the planetary bond albedo is 0.3 \cite[e.g.,][]{2014ApJ...789L..20D}. Finally, we maintain the heating efficiency ($\epsilon$) at a constant value of 0.3, following the assumption from \citet{2016ApJ...829...63S}. 

%%%%%%%%%%%%%%%%%%%%
\begin{figure}
\includegraphics[width=18cm]{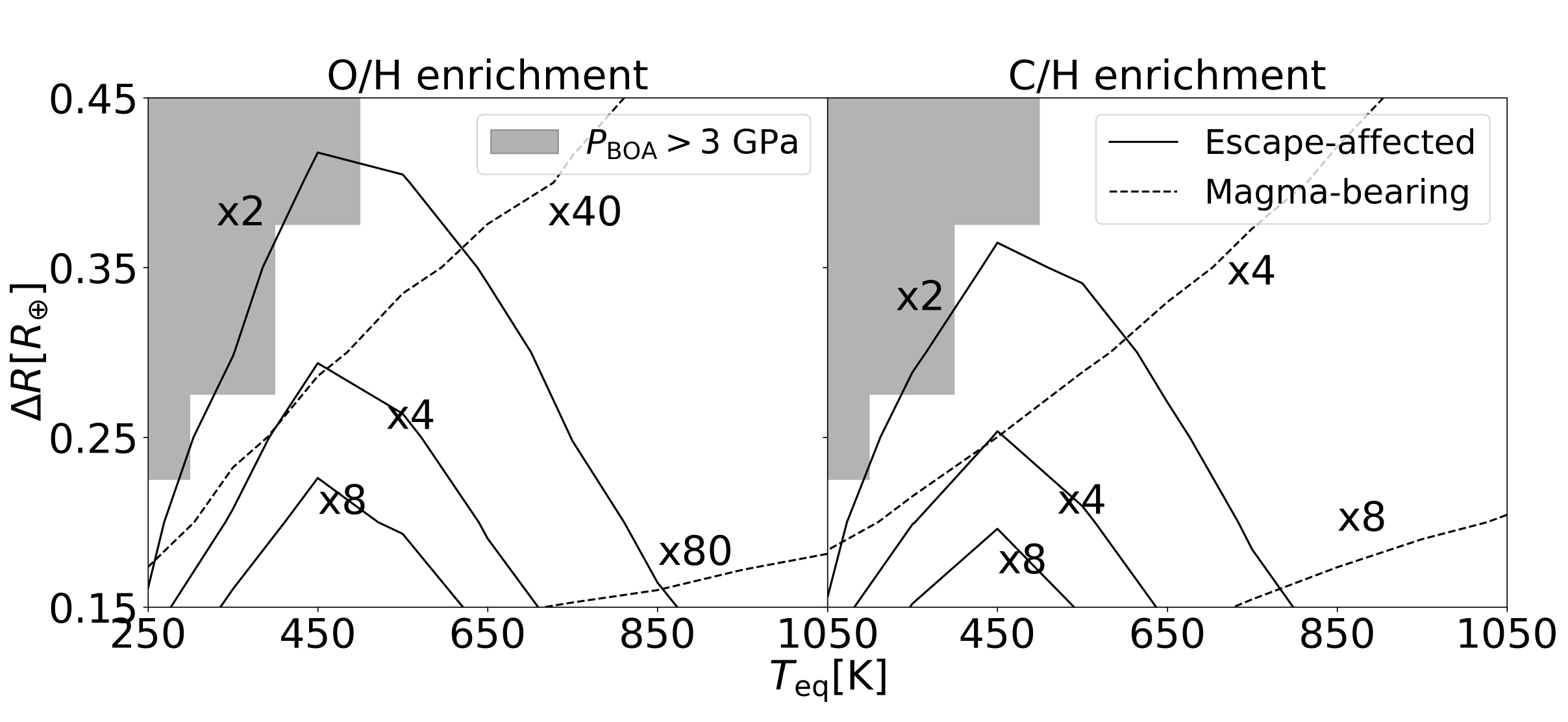} 
\caption{The enrichment in atmospheric O/H (left panel) and C/H ratio (right panel) relative to the initial nebula value due to species-dependent atmospheric escape (solid contour) and due to the interaction with magma (dashed contour). The x-axis denotes the planetary equilibrium temperature while the y-axis represents the atmospheric thickness. The black region indicates the parameter space where the BOA pressure is larger than the validated range ($\lesssim$3~GPa) for the calculation of magma-bearing cases.}
\label{fig:escape}
\end{figure} 
%%%%%%%%%%%%%%%%%%%%

The solid lines in Figure \ref{fig:escape} show the atmospheric O/H and C/H enrichment of 6$\rm M_{\oplus}$ sub-Neptunes due to the atmospheric escape as a function of atmospheric thickness and equilibrium temperature. The dashed lines represent the enrichment due to interaction with the magma ocean (without the effect of atmospheric loss), where the $N_{\rm O(Fe)}^{\rm before}$/$N_{\rm Fe}$ ratio is set to 1.0. Data with $\Delta R=0.1R_{\rm \oplus}$ are excluded from the figure because magma cannot exist if $N_{\rm O(Fe)}^{\rm before}$/$N_{\rm Fe} = 1$ under such thin atmospheres due to low BOA temperature. The black region in Figure \ref{fig:escape} indicates the region where the surface pressure exceeds 3~GPa, which is beyond the pressure range validated by experiments. We excluded this non-validated region from our calculations. The enrichment due to atmospheric escape is most pronounced when $T_{\rm eq} \sim 450$~K, as a compromise between the amount of atmospheric loss (higher at higher equilibrium temperature) and the efficiency of the selective escape (higher at lower equilibrium temperature). As expected, thinner atmospheres exhibit a stronger atmospheric escape effect. 

We find that in the broad parameter space shown here, the effect of a magma ocean on O/H enrichment is dominant over that of atmospheric escape. Additionally, while magma does not effectively enrich atmospheric C/H, atmospheric escape enriches C/H to the same level as O/H because of the similar mass of C and O. Therefore, large O/H with small C/O (relative to the nebula values) would be safely attributed to the effect of the interaction between the atmosphere and magma, rather than the atmospheric escape. Escape can be accelerated by ionization \cite[e.g.,][]{2023AJ....166...89D} compared to the neutral-neutral pair, in which case the enrichment of C/H and O/H due to escape is likely to be smaller. However, the effect on the C/O ratio is limited.

\subsection{Dependence on the atmospheric structure: Comparison to the radiative-convective profile}
\label{ss:discussion_TP}

Our fiducial models assumed a simplified atmospheric T-P profile consisting of an isothermal stratosphere and adiabatic troposphere, which simplifies the analytic interpretation. The temperature profiles of real planets can deviate from this simple profile, and the relation between the observable parameters and the condition at the BOA can be different from our nominal models. To evaluate this effect, we additionally calculated temperature profiles in one-dimensional radiative-convective equilibrium using the open-source code HELIOS \citep{2017AJ....153...56M,2019AJ....157..170M}. 
In this subsection, we present the calculation set-up and the results, showing that the results with our nominal assumptions exhibit similar planetary parameter dependencies as discussed in Section \ref{ss:atmospheric composition_planetaryparamdepen} but with a lower O/H-C/H ratio and stronger C/O depletion.

The calculation of radiative-convective equilibrium profiles with HELIOS requires (1) the assumption for the atmospheric composition, (2) the opacities of considered atmospheric species, and (3) the adiabtatic T-P profile for the convective regions. 
For (1), we consider H$_2$, He, H$_2$O, Na, and K. 
We ignore the presence of C-bearing species because they hardly affect the opacity of the atmosphere and the mean molecular weight. We should emphasize that the T-P profiles can be affected by the species not considered in our chemistry model. In particular, alkali metals such as Na and K have significant opacities at the optical wavelengths where \HH, \He, and \HHO\ are largely transparent and thus substantially alter the T-P profile. The extensive survey of the range of the T-P profile will be the scope of future papers. In this paper, we simply set the ratio of He/H, Na/H, and K/H to be the values of the solar-composition atmosphere in chemical equilibrium at the given temperature, while the \HHO\ mixing ratio is varied. 

As for (2), the k-coefficient tables up to 10$^5$ bar were calculated with HELIOS-K \citep{2015ApJ...808..182G, 2021ApJS..253...30G}. The opacity tables for the collision-induced absorption, H$_2$-H$_2$ and H$_2$-He, were created with HITRAN \citep{2022JQSRT.27707949G}, and that for \HHO\ was based on the ExoMOL database \citep{2018MNRAS.480.2597P}. We extrapolate the opacity tables for Na and K from the tables provided in the default dataset up to 1000 bar, assuming that the opacity is proportional to the pressure.

Regarding (3), the adiabatic $\rm T$-$\rm P$ profile, specifically $d \log T/d \log P$, was computed in a way consistent with our default model taking account of the temperature dependence of $C_p$ for \HH, \HHO, and \He. The $\rm T$-$\rm P$ profile deeper than $10^5$ bar is simply extrapolated using the adiabatic lapse rate at the BOA. 

Other major scalar input parameters include the internal temperature, which is assumed to be 50~K, similar to the value of Neptune \citep{1991JGR....9618921P}, the BOA pressure, which is set to $10^5$~bar, and the stellar parameters where the radius is 0.45 $R_{\odot }$ and the effective temperature is 3500~K. 
The surface gravity in the radiative transfer routine is fixed to 20~${\rm m/s^2}$, close to but smaller than the surface gravity at the BOA for our nominal case (23.4 ${\rm m/s^2}$); the variation of surface gravity in the range of 10--23.4 ${\rm m/s^2}$ (corresponding to $0 < \Delta R < 0.8 R_{\oplus }$) has only minor effects in the $\rm T$-$\rm P$ profile. 

\begin{figure}
\includegraphics[width=18cm]{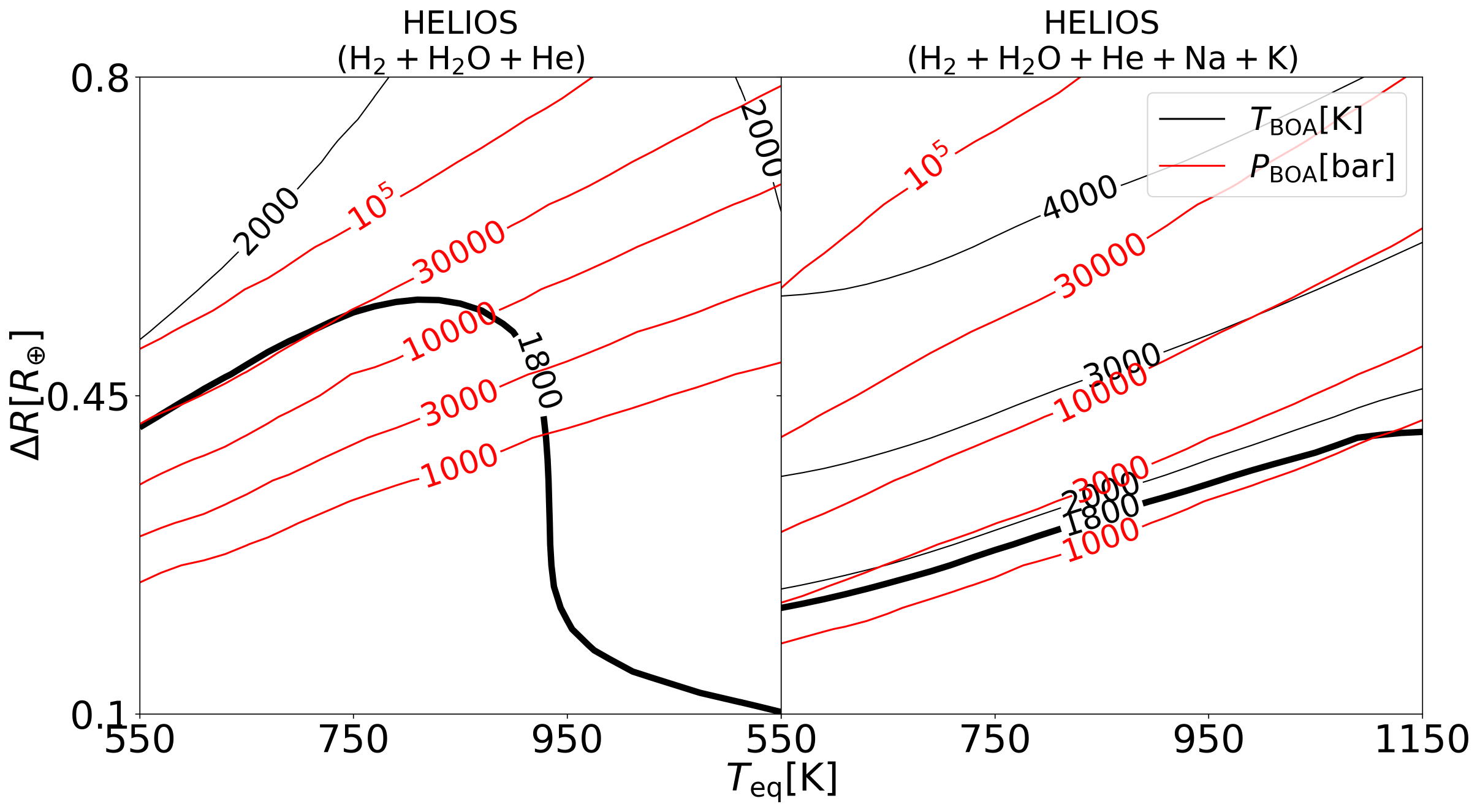} 
\caption{The relationship among $P_{\rm BOA}$ (red contour), $T_{\rm BOA}$ (black contour), $\Delta R$, and $T_{\rm eq}$ in the cases of radiative-convective equilibrium atmospheric profile. The radiative-convective equilibrium profile is further divided into the profile without Na and K (left) and including them (right). The x-axis denotes the planetary irradiation (equilibrium) temperature, while the y-axis represents the atmospheric thickness. The atmospheric composition is assumed to be \HH\ (82\%), He (17\%), \HHO (1\%), Na (solar metallicity), and K (solar metallicity).}
\label{fig:HELIOS}
\end{figure}

After computing the radiative-convective $\rm T$-$\rm P$ profile, we use these profiles to find the surface pressure and temperature given $\Delta R$, now taking account of the altitude dependence of surface gravity.
Figure \ref{fig:HELIOS} shows the relation among $\Delta R$, $T_{\rm eq}$, $P_{\rm BOA}$, and $T_{\rm BOA}$ based on the radiative-convective calculation of HELIOS (see Figure \ref{fig:surfaceconditions} for comparison). Here, $T_{\rm eq}$ is estimated from the irradation flux $F_{\rm irr}$ by $T_{\rm eq} \sim ( F_{\rm irr}/(4\sigma_{\rm SB}))^{1/4}$, where $\sigma _{\rm SB} $ is the Stefan-Boltzmann coefficient.

%%%%%%%%%%%%%%%%%%%%
%%%%%%%%%%%%%%%%%%%%

Notably, the HELIOS results indicate a much smaller $T_{\rm BOA}$ than $T_{\rm BOA}$ from the T-P profile of our model if effective opacity sources like Na and K are excluded. On the other hand, the inclusion of Na and K maintains relatively higher $T_{\rm BOA}$, which is slightly lower than or comparable to the $T_{\rm BOA}$ in the case of our model assumption. 
Consequently, the range of planetary parameters (atmospheric thickness and equilibrium temperature) that ensures the presence of magma would be much narrower compared to the result with our nominal model setting if Na and K cannot contribute as effective opacity sources. 
However, it is essential to note that the effect of different values for $T_{\rm BOA}$ on the trend and range of atmospheric composition is limited, as discussed in Section \ref{ss:atmospheric composition}. 

Conversely, the $P_{\rm BOA}$ of the HELIOS result is larger than that of our model. Consequently, atmospheric composition from the HELIOS profile is expected to have lower O/H and C/H ratios under the same $N_{\rm O(Fe)}^{\rm before}$/$N_{\rm Fe}$, as illustrated in Figure \ref{fig:atmos_Oini_linear_surfacedepen}. In addition, this increased pressure effect will further lower the C/O ratio with a given O/H ratio, reinforcing our result that magma leads to a depleted atmospheric C/O ratio (see Figure \ref{fig:OHCOrelation}). 

\subsection{Limitations}
\label{ss:Limitation}

There are mechanisms not covered in our model that are expected to suppress the C/O ratio further by either decreasing the amount of C or increasing the amount of O. The precipitation of C-bearing species into graphite or diamond can influence carbon partitioning, potentially reducing the atmospheric carbon content (e.g., \citealt{2012E&PSL.341...48H} and \citealt{2024arXiv240509284M}). Mechanisms that increase the amount of O include the precipitation of iron droplets. Contrary to our assumption that iron droplets are suspended in magma, weaker magma convection would precipitate them, resulting in an oxidized magma. The partitioning of silicon between magma and the iron core ($\rm 2Fe_{metal} + SiO_{2} \longleftrightarrow 2FeO + Si_{metal}$) may also introduce additional FeO into the mantle, leading to a more oxidized state than would otherwise be expected from the same initial rock composition. These mechanisms only reinforce our conclusion that magma can deplete the atmospheric C/O ratio. 

In addition to the oxidation of magma through silicon partitioning mentioned above, the role of the innermost iron core as a reservoir for hydrogen, carbon, and oxygen through efficient partitioning is widely recognized \cite[e.g.,][]{2018GeoRL..45.6042D, 2020PNAS..117.8743F, 2020NatGe..13..453L, 2022PSJ.....3..127S, 2023FrEaS..1159412S, 2023PSJ.....4...30H, 2023PNAS..12009786K}. Given that the iron core constitutes approximately 34\% of the total planetary mass, volatile partitioning to the iron core can significantly impact atmospheric composition. However, these partitioning coefficients are sensitive to pressure, temperature, and silicate composition, emphasizing the importance of understanding the core formation process for accurate characterization. Therefore, a detailed discussion of the impact of the iron core on atmospheric composition remains in our future scope.

There are admittedly uncertainties in our toy model of the solidification process. 
In particular, approximating the solidification as a two-step process likely overestimates the final atmospheric O, as the gradual oxidization of the magma-atmosphere system during solidification would leave more oxidized iron in the solidified rock. 
Also, the trapping of \HHO\ in the solidified layer \citep[e.g.,][]{2023SSRv..219...51S} and the separation of magma into shallow surface and deep basal magma layers due to mid-crystallization \cite[e.g.,][]{2019E&PSL.516..202C}, which are ignored here, would suppress the oxidation of the magma-atmosphere system. 
While our conclusion that the solidification further oxidizes the atmosphere is likely to be valid, the quantitative assessment of the effect of the above-mentioned processes requires further modeling.

The exclusion of rocky vapor in our model introduces uncertainties in the results, particularly under high temperatures. For example, rocky vapor can influence atmospheric composition through reactions such as the formation of silane: $\rm SiO + 3H_{2} \longleftrightarrow SiH_{4} + H_{2}O$ (e.g., \citealt{2023MNRAS.524..981M}). Furthermore, the atmospheric structure can be affected by rocky vapor (e.g., \citealt{2022MNRAS.514.6025M}). Investigating the impact of rocky vapor on the internal chemical reaction system and upper atmospheric composition will be our future scope.

We anticipate that future geochemical studies will provide important implications, especially for the parameter ranges not covered in this study due to a lack of validated data. For instance, the dependence of solubility on melt composition has not been extensively investigated, except for a few species like \HH, \COO, and \HHO\ \cite[e.g.,][]{2012E&PSL.345...38H, 2022CoMP..177...40A}. This is why we do not include a comprehensive analysis of melt composition (except for Fe content) in this study. If we were to infer the effect of melt composition on \HH\ solubility (which has been discussed for three cases: Andesite, Basalt, and Peridotite; see \citealt{2012E&PSL.345...38H}), peridotite melt could reduce \HH\ solubility by about a factor of three, further depleting atmospheric C/H and C/O ratios. The melt composition effects on the other two gases (\HHO\ and \COO) are likely limited since the dependence of \HHO\ is less than factor of two for $f{\rm {H_2O}} \leq 2000$~bar \citep[e.g.,][]{2022CoMP..177...40A} and \COO\ is the most minor dissolving species in our model (see also Figure \ref{fig:atmos_magma_species_Oini}). The dependence of the solubility on the temperature is not well constrained from data either. Given that many sub-Neptunes likely have high-temperature and high-pressure surface conditions, the solubility data under extreme temperature-pressure environments will be invaluable for future modeling to properly interpret the sub-Neptune data.

\section{Conclusion}

Motivated by the bulk composition degeneracy of sub-Neptunes, this study aimed to investigate the range of atmospheric H-C-O composition in magma-bearing sub-Neptunes formed through the nebula gas accretion on a dry, rocky core. We clarify the trend in the atmospheric composition and their dependence on magma properties (effective amount, Fe fraction, and initial Fe speciation), observable planetary parameters (planetary mass, equilibrium temperature, and radius), and an additional parameter for determining atmospheric structure (Radiative-Convective Boundary), using a chemical equilibrium model between the atmosphere and magma core. Furthermore, We discuss the effect of magma solidification on atmospheric composition. Our key findings are summarized as follows:

1. The atmospheric O/H ratio (or equivalently $x_{\rm H_{2}O}$) after the reaction between the accreted volatile and magma is a function of the redox state of the magma (which we parameterize by the number of iron-bound oxygen before the reaction, $N_{\rm O(Fe)}^{\rm before}$/$N_{\rm Fe}$) and the surface pressure ($\rm P_{\rm BOA}$). The atmospheric \HHO\ fraction is proportional to $N_{\rm O(Fe)}^{\rm before}$/$N_{\rm Fe}$ at low pressure ($P_{\rm BOA}\lesssim 10000\ \rm bar$) and $(N_{\rm O(Fe)}^{\rm before}/N_{\rm Fe})^{\frac{1}{0.74}}$$P_{\rm BOA}^{-1}$ at high pressure ($P_{\rm BOA} \gtrsim 10000\ \rm bar$). The atmospheric \HHO\ fraction ranges from a few percent to a few tens of percent (only for thin atmospheres, $P_{\rm BOA} << 10000\ \rm bar$) unless the initial rock is substantially reduced, with more than about 90\% of Fe present in the form of pure iron.

2. Despite the higher solubility of \HH\ and \HHO\ than C-bearing species, the C/H ratio experiences a modest increase, typically two to three times higher than the accreted (nebula) value. Only in highly oxidized thin-atmosphere scenarios does the C/H ratio exhibit enrichment of several tens of times.

3. These two trends can be summarized in a simple relation between the atmospheric C/O ratio and atmospheric O/H ratio (or equivalently $x_{\rm H_{2}O}$) under the existence of magma. The atmospheric C/O ratio is proportional to $x_{\rm H_{2}O}^{-1}$ under a reduced atmosphere and $x_{\rm H_{2}O}^{-0.26}$ under an oxidized atmosphere. Thus, the atmospheric C/O ratio decreases to below 10\% of the nebula value under an oxidized atmosphere with larger than a few percent of \HHO\ fraction. 

4. A higher magma Fe fraction leads to a more oxidized atmosphere under constant $N_{\rm O(Fe)}^{\rm before}$/$N_{\rm Fe}$, as it increases the oxygen available for atmospheric-magma reactions. However, the Fe fraction does not alter the O/H-C/O relationship itself, maintaining a similar range of C/O ratios regardless of the Fe fraction.

5. A smaller amount of the reactive magma causes the atmospheric composition to approach that of the accreted volatiles. A significant shift in the O/H ratio trend is observed when the reactive magma fraction falls below $\sim $10\%, assuming $M_{\rm p} = 6 M_{\oplus}$, $x_{\rm H_{2}O} = 0.01$, and $P_{\rm BOA}=10000$~bar. Moreover, the quantity of reactive magma influences the O/H-C/O trend, where a smaller volume of the reactive magma leads to a smaller C/O ratio for a given O/H ratio.

6. A thicker atmosphere with low equilibrium temperature, high radiative-convective boundary pressure, and heavier planetary core results in lower atmospheric O/H and C/H ratios due to the larger surface pressure, again not changing the O/H-C/O relation. 

7. Solidification of the magma further increases the atmospheric O/H ratio and decreases the atmospheric C/O ratio, potentially by one or two orders of magnitude, through the extraction of \HHO\ to the magma-atmosphere system.

Our results suggest that the atmospheric O/H - C/O ratio relation and the resulting depletion of the atmospheric C/O ratio to below 10\% of the nebula value under an oxidized atmosphere can be evidence of magma-atmospheric interaction {as well as no substantial additional supply of C. Further constraints on the volatile composition of icy solids from which sub-Neptunes could have formed would provide clearer insights into the distinction between icy sub-Neptunes from those formed through the nebula gas accretion onto the rocky core.

\begin{acknowledgments}
The authors deeply thank E. S. Kite for fruitful comments and H. Kuwahara and T. Tsuchiya for the discussions on the material properties at high pressure. 
YI was supported by JSPS KAKENHI grant 22K14090. 
\end{acknowledgments}

\appendix

\section{The differences in results based on the extrapolation method for the equilibrium constant.}
\label{ap:extrapolation}

In this section, we present the results obtained with a different extrapolation method for equilibrium constants. As explained in section \ref{ss:model_redox_chemistry}, we extrapolate the equilibrium coefficient equations for the IW reaction and Ferric-Ferrous equilibrium up to 4000 K even though the validation range for Ferric-Ferrous equilibrium ends at 2200 K. Figure \ref{fig:Mainres_forappendix2} shows the results where we simply use the equilibrium constants at the boundary value of 2200 K if the temperature of the system exceeds 2200 K. The two results (Figure \ref{fig:atmos_Oini_linear_surfacedepen} and Figure \ref{fig:Mainres_forappendix2}) show that the boundary value extrapolation leads to a slightly more oxidized atmospheric composition, consistent with expectations from Figure \ref{fig:atmos_Oini_linear_surfacedepen}. However, the difference is minute and does not affect the main results of this study.

%%%%%%%%%%%%%%%%%%%%
\begin{figure*}
\centering
\includegraphics[width=18cm]{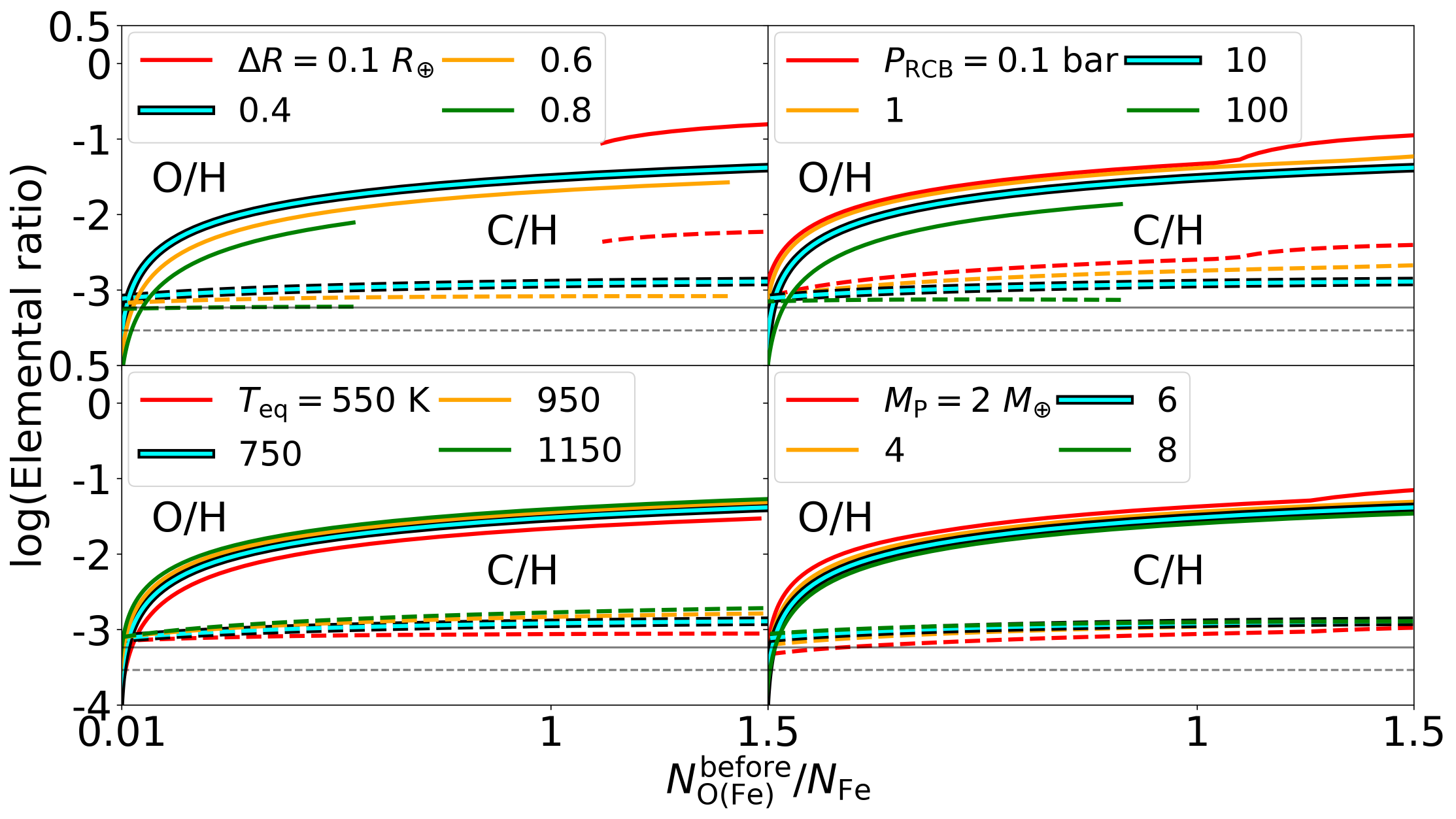}
\caption{The same figure as Figure \ref{fig:atmos_Oini_linear_surfacedepen}, but generated using results obtained with a different extrapolation method. Note the negligible difference from Figure \ref{fig:atmos_Oini_linear_surfacedepen}.}
\label{fig:Mainres_forappendix2}
\end{figure*}
%%%%%%%%%%%%%%%%%%%%

\bibliography{ref}

\end{document}